\documentclass[reprint,amssymb, amsmath, aps, superscriptaddress, showpacs, footinbib,  prb]{revtex4-1}
\pdfoutput=1
\usepackage{graphicx}

\newcommand{\eff}{\text{eff}}
\newcommand{\AFM}{\text{AFM}}
\newcommand{\FM}{\text{FM}}

\newcommand{\mg}{\text{mag}}

\newcommand{\sub}{\text{sub}}
\newcommand{\iso}{\text{iso}}
\newcommand{\str}{\text{str}}

\newcommand{\navpof}{Na$_{1.5}$VOPO$_4$F$_{0.5}$}

\begin{document}

\title{Phase separation and frustrated square lattice magnetism of Na$_{1.5}$VOPO$_4$F$_{0.5}$}

\author{A.~A.~Tsirlin}
\email{altsirlin@gmail.com}
\affiliation{Max Planck Institute for Chemical Physics of Solids, N\"{o}thnitzer
Str. 40, 01187 Dresden, Germany}

\author{R.~Nath}
\affiliation{Max Planck Institute for Chemical Physics of Solids, N\"{o}thnitzer
Str. 40, 01187 Dresden, Germany}
\affiliation{Ames Laboratory and Department of Physics and Astronomy, Iowa State University, Ames, Iowa 50011 USA}
\affiliation{School of Physics, Indian Institute of Science Education and Research,
Trivandrum-695016 Kerala, India}

\author{A.~M.~Abakumov}
\affiliation{EMAT, University of Antwerp, Groenenborgerlaan 171, B-2020 Antwerp, Belgium}

\author{Y.~Furukawa}
\author{D.~C.~Johnston}
\affiliation{Ames Laboratory and Department of Physics and Astronomy, Iowa State University, Ames, Iowa 50011 USA}

\author{M.~Hemmida}
\author{H.-A.~{Krug von Nidda}}
\author{A.~Loidl}
\affiliation{Experimental Physics V, Center for Electronic Correlations and Magnetism, University of Augsburg, 86135 Augsburg, Germany}

\author{C.~Geibel}
\author{H.~Rosner}
\email{Helge.Rosner@cpfs.mpg.de}
\affiliation{Max Planck Institute for Chemical Physics of Solids, N\"{o}thnitzer
Str. 40, 01187 Dresden, Germany}

\begin{abstract}
Crystal structure, electronic structure, and magnetic behavior of the spin-$\frac12$ quantum magnet Na$_{1.5}$VOPO$_4$F$_{0.5}$ are reported. The disorder of Na atoms leads to a sequence of structural phase transitions revealed by synchrotron x-ray powder diffraction and electron diffraction. The high-temperature second-order $\alpha\leftrightarrow\beta$ transition at 500~K is of the order-disorder type, whereas the low-temperature $\beta\leftrightarrow\gamma+\gamma'$ transition around 250~K is of the first order and leads to a phase separation toward the polymorphs with long-range ($\gamma$) and short-range ($\gamma'$) order of Na. Despite the complex structural changes, the magnetic behavior of \navpof\ probed by magnetic susceptibility, heat capacity, and electron spin resonance measurements is well described by the regular frustrated square lattice model of the high-temperature $\alpha$-polymorph. The averaged nearest-neighbor and next-nearest-neighbor couplings are $\bar J_1\simeq -3.7$~K and $\bar J_2\simeq 6.6$~K, respectively. Nuclear magnetic resonance further reveals the long-range ordering below $T_N=2.6$~K in low magnetic fields. Although the experimental data are consistent with the simplified square-lattice description, band structure calculations suggest that the ordering of Na atoms introduces a large number of inequivalent exchange couplings that split the square lattice into plaquettes. Additionally, the direct connection between the vanadium polyhedra induces an unusually strong interlayer coupling having effect on the transition entropy and the transition anomaly in the specific heat. Peculiar features of the low-temperature crystal structure and the relation to isostructural materials suggest \navpof\ as a parent compound for the experimental study of tetramerized square lattices as well as frustrated square lattices with different values of spin.
\end{abstract}

\pacs{75.50.Ee, 75.30.Et, 75.10.Jm, 61.66.Fn}
\maketitle

\section{Introduction}
\label{sec:intro}
Among a broad family of low-dimensional and frustrated magnets, systems based on the frustrated square lattice (FSL) model enjoy special attention from theory and experiment. The model entails competing exchange couplings along sides ($J_1$) and diagonals ($J_2$) of the square and is mostly studied for purely Heisenberg Hamiltonian and spin-$\frac12$. Extensive theoretical work\cite{misguich,shannon2004} convincingly established three ordered ground states that emerge for different values of the frustration ratio $\alpha=J_2/J_1$. The N\'eel and columnar antiferromagnetic (AFM) states are separated by a critical spin-liquid region around the quantum critical point at $\alpha=\frac12$ (Refs.~\onlinecite{darradi2008,isaev2009,reuther2010,richter2010a}). At $\alpha=-\frac12$, the columnar AFM phase should border the ferromagnetic (FM) phase. However, the precise nature of this boundary remains controversial. While an earlier report proposed a nematic phase separating the regions of FM and columnar AFM ground states,\cite{shannon2006} Richter \textit{et al.}\cite{richter2010} demonstrated the single abrupt transition without any intermediate phases around $\alpha=-\frac12$. 

Experimentally, the FSL-type magnetic behavior has been observed in V$^{+4}$ compounds with [VOXO$_4$] layers comprising VO$_5$ square pyramids and non-magnetic XO$_4$ tetrahedra\cite{melzi2000,carretta2002,kaul2004} (the only known exception is PbVO$_3$ with magnetic layers formed by VO$_5$ pyramids exclusively\cite{tsirlin2008,*oka2008}). If X is a main group ($p$) cation, the leading coupling is AFM $J_2$, whereas $J_1$ is usually weaker and can be either FM (X = P) or AFM (X = Si, Ge), see also Table~\ref{tab:comparison}.\cite{rosner2002,kaul2004,tsirlin2009} Recent efforts in crystal growth\cite{kaul-thesis} and experimental investigation\cite{melzi2001,carretta2002a,carretta2009,high-field,bossoni2010} of such compounds established the columnar AFM ground state\cite{bombardi2004,skoulatos2009,nath2009} and revealed simple trends in thermodynamic properties,\cite{kaul-thesis,nath2008,tsirlin2010} in line with theoretical predictions.\cite{shannon2004} A thorough comparison between experiment and theory, however, spots certain discrepancies. For example, the sublattice magnetization is gradually reduced from 0.6~$\mu_B$ in Li$_2$VOSiO$_4$ ($\alpha\simeq 10$) to 0.5~$\mu_B$ in Pb$_2$VO(PO$_4)_2$ ($\alpha\simeq -2$) and 0.4~$\mu_B$ in SrZnVO(PO$_4)_2$ ($\alpha\simeq -1$).\cite{bombardi2004,skoulatos2009} By contrast, theory predicts a nearly constant sublattice magnetization of 0.6~$\mu_B$ in this range of $\alpha$ (Ref.~\onlinecite{richter2010}). The discrepancy may arise from the spatial anisotropy of the FSL due to the low crystallographic symmetry of Pb$_2$VO(PO$_4)_2$ and SrZnVO(PO$_4)_2$. In fact, none of the reported FSL-type phosphates, $AA'$VO(PO$_4)_2$ ($AA'$ = Pb$_2$, PbZn, SrZn, BaZn, BaCd) with FM $J_1$, are tetragonal, thereby the effects of the spatial anisotropy are expected.\cite{tsirlin2009}

\begin{table*}[!ht]
\begin{minipage}{14cm}
\caption{\label{tab:parameters}
Lattice parameters, space groups, and refinement residuals for different polymorphs of \navpof. 
}
\begin{ruledtabular}
\begin{tabular}{cccccc}
  Polymorph & Temperature & $a$         & $c$         & Space group & $R_I/R_p$   \\
            & (K)         & (\r A)      & (\r A)      &             &             \\
  $\alpha$  & 560         & 6.39563(1)  & 10.65908(2) & $I4/mmm$    & 0.031/0.109 \\
  $\beta$   & 298         & 9.03051(2)  & 10.62002(3) & $P4_2/mnm$  & 0.035/0.103 \\
  $\gamma$  & 150         & 12.76716(2) & 10.57370(4) & $P4_2/mbc$  & 0.033/0.073 \\
  $\gamma'$ & 150         & 6.37996(4)  & 10.5910(1)  & $I4/mmm$    & 0.022/0.073 \\
\end{tabular}
\end{ruledtabular}
\end{minipage}
\end{table*}
To facilitate the experimental verification of theoretical results for the FSL model, tetragonal systems with a perfect square lattice of magnetic atoms are required. Motivated by this challenge, we explored the structure and properties of \navpof. This compound was prepared in 2002 by a hydrothermal method. Massa \textit{et al.}\cite{massa2002} found a tetragonal crystal structure with the three-dimensional (3D) VOPO$_4$F$_{0.5}$ framework formed by the FSL-type VOPO$_4$ layers (Fig.~\ref{fig:structure}). Sauvage \textit{et al.}\cite{sauvage2006} claimed to prepare the same compound by a high-temperature annealing in air, and reported a very similar structure refinement. Since Na atoms in \navpof\ and related systems are readily deintercalated and exchanged with Li,\cite{sauvage2006,gover2006} the frustration ratio $\alpha$ could be tuned, making \navpof\ an appealing FSL system. To explore this possibility, we performed a comprehensive study of the parent compound \navpof. 

The outline of the paper is as follows. We list experimental and computational procedures in Sec.~\ref{sec:methods}, and proceed to details of the crystal structure in Sec.~\ref{sec:structure}. Further on, we perform thermodynamic and magnetic resonance measurements (Sec.~\ref{sec:experiment}) and band structure calculations (Sec.~\ref{sec:band}) to evaluate the low-temperature magnetic behavior and the microscopic magnetic model. The comparison of \navpof\ to known FSL compounds and other structural analogs is given in Sec.~\ref{sec:discussion} followed by a summary and outlook.
\section{Methods}
\label{sec:methods}
Powder samples of \navpof\ were prepared by a solid-state reaction of Na$_4$P$_2$O$_7$, VO$_2$, and NaF in an evacuated and sealed silica tube at 700~$^{\circ}$C for 24 hours. The stoichiometric mixtures of reactants were pelletized, placed into corundum crucibles, and covered with lids to avoid the reaction between silica and NaF. Na$_4$P$_2$O$_7$ was obtained by the decomposition of Na$_2$HPO$_4$ in air at 400~$^{\circ}$C. The bluish-green powders of \navpof\ were single-phase, as confirmed by laboratory x-ray diffraction (XRD) measured with Huber G670 Guinier camera (CuK$_{\alpha1}$ radiation, $2\theta=3-100^{\circ}$ angle range, image-plate detector). 

High-resolution XRD data for structure refinement were collected in the $150-560$~K temperature range at the ID31 beamline of European Synchrotron Radiation Facility (ESRF) with a constant wavelength of about 0.4~\r A. The signal was measured by eight scintillation detectors, each preceded by a Si (111) analyzer crystal, in the angle range $2\theta=1-40$~deg. The powder sample was contained in a thin-walled borosilicate glass capillary with an external diameter of 0.5~mm. The sample was cooled below room temperature in a He-flow cryostat and heated above room temperature with a hot-air blower. To achieve good statistics and to avoid the effects of the preferred orientation, the capillary was spun during the experiment. The JANA2006 program was used for the structure refinement.\cite{jana2006} Symmetry changes at the structural phase transitions were analyzed with ISODISPLACE program.\cite{isodisplace}

The samples for an electron diffraction (ED) study were prepared by crushing the powder in ethanol and depositing the suspension on a holey carbon grid. ED patterns were taken at room temperature (RT) and at 100~K using a Philips CM20 microscope equipped with a Gatan cooling holder.

The differential scanning calorimetry (DSC) measurement was performed with a Perkin Elmer DSC 8500 instrument in the temperature range $120-800$~K in argon atmosphere with a heating/cooling rate of 10~K/min. The powder sample of \navpof\ was placed into a corundum crucible.

The magnetic susceptibility of \navpof\ was measured with an MPMS SQUID magnetometer in the temperature range $2-380$~K in applied fields up to 5~T. Heat capacity measurements on a pressed pellet were performed by relaxation technique with a Quantum Design PPMS instrument in the temperature range $0.45-200$~K and in fields up to 11~T. The data below 1.8~K were collected with a $^3$He insert.

The electron spin resonance (ESR) measurements were performed at X-band frequency ($\nu=9.36$~GHz) on a Bruker ELEXSYS E500-CW spectrometer, equipped with a continuous He-gas flow cryostat (Oxford Instruments) operating in the temperature range $4.2-300$~K. The polycrystalline powder sample was fixed in a quartz tube with paraffin and mounted in the center of the microwave cavity. The field derivative of the microwave--absorption signal was detected as a function of the static magnetic field due to the lock-in technique with field modulation. Resonance absorption occurs when the incident microwave energy matches the energy of magnetic dipolar transitions between the electronic Zeeman levels.

The nuclear magnetic resonance (NMR) measurements were carried out using pulsed NMR techniques on $^{31}$P (nuclear spin $I=\frac12$ and gyromagnetic ratio $\gamma_N/2\pi = 17.237$ MHz/T) nuclei in the temperature range $1.5-300$~K. We did the measurements at radio frequencies of $79$~MHz and $8.62$~MHz, which correspond to applied fields of about $4.583$~T and $0.5$~T, respectively. The spectra were obtained either by sweeping the field or by doing the Fourier transform. The NMR shift $K=(H_{\rm ref}-H)/H$ was determined by measuring the resonance field of the sample ($H$) with respect to a nonmagnetic reference H$_{3}$PO$_{4}$ (resonance field $H_{\rm ref}$). The $^{31}$P spin-lattice relaxation rate $1/T_1$ was measured by the conventional single saturation pulse method.

Scalar-relativistic band-structure calculations for \navpof\ were performed within the framework of density functional theory (DFT) using the basis set of local orbitals (\texttt{FPLO9.01-35} code).\cite{fplo} We applied the local density approximation (LDA) with the exchange-correlation potential by Perdew and Wang\cite{pw92} and a $k$ mesh of 216~points in the first Brillouin zone (40~points in the irreducible wedge). Exchange couplings were evaluated by mapping V $3d$ bands onto a multi-orbital Hubbard model (see Sec.~\ref{sec:band} for further details).
\section{Crystal structure}
\label{sec:structure}
\subsection{High-temperature $\alpha$-polymorph}
According to RT studies of Refs.~\onlinecite{massa2002,sauvage2006}, \navpof\ has a body-centered tetragonal unit cell with $a_{\sub}\simeq 6.38$~\r A and $c\simeq 10.62$~\r A. Our RT XRD pattern is largely consistent with this unit cell, although 10 weak reflections remained unindexed. These reflections can be assigned to the $\sqrt2\,a_{\sub}\times\sqrt2\, a_{\sub}\times c$ supercell and evidence the superstructure formation at RT. The superstructure reflections in \navpof\ disappeared upon heating. Above 500~K, the patterns could be fully indexed on the $a_{\sub}\times a_{\sub}\times c$ body-centered tetragonal unit cell that we further refer as $\alpha$-modification, in contrast to $\beta$-modification at RT (Table~\ref{tab:parameters}).

\begin{figure}
\includegraphics{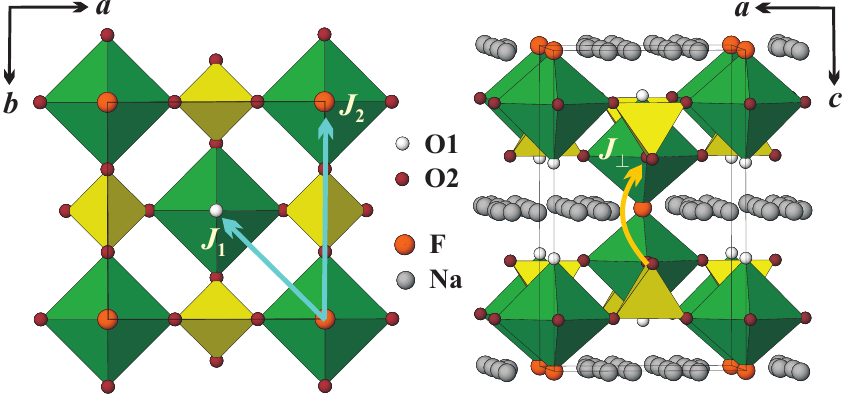}
\caption{\label{fig:structure}
(Color online) Crystal structure of \mbox{$\alpha$-\navpof:} the layers in the $ab$ plane (left) and the projection along $c$ (right). The layers have an ideal FSL geometry with the couplings $J_1$ and $J_2$ (left) and connect along $c$ via the F atoms (right). Disordered Na atoms occupy the voids of the resulting framework (right).
}
\end{figure}
The structure refinement\cite{supplement} for the $\alpha$-phase (Table~\ref{tab:alpha}) conforms to the atomic positions given in Ref.~\onlinecite{massa2002}. We find a 3D framework of corner-sharing VO$_5$F octahedra and PO$_4$ tetrahedra (Fig.~\ref{fig:structure}). Vanadium atoms form one short bond to the O1 atom (along the $c$ direction), four longer bonds to oxygens in the $ab$ plane (the O2 position), and one long bond to the fluorine atom (also along $c$), see Table~\ref{tab:distances}. The octahedra are linked via the PO$_4$ tetrahedra in the $ab$ plane, whereas the fluorine atoms connect the octahedra into pairs along $c$. Na atoms occupy the voids of the resulting framework.

At this point, the question concerning the arrangement of O and F atoms may arise. Since atomic numbers of these elements differ by unity, it is hard to distinguish between O and F using XRD, especially in a powder experiment. The single-crystal refinement\cite{massa2002} proposed the complete ordering of the O and F atoms. The ordered arrangement of these atoms is also supported by empirical arguments. The O1 position corresponds to the short V--O bond which is typical for oxygen and uncommon for fluorine, e.g., in oxyfluorides.\cite{[{For example: }][{}]aldous2007} Further on, the O2 atoms in the $ab$ plane are parts of rigid PO$_4$ tetrahedra and can not mix with fluorine. Therefore, fluorine is left to its $2a$ position with two long bonds to vanadium. Note also that our ESR and NMR experiments (Sec.~\ref{sec:experiment}) suggest unique (or very similar) positions for vanadium and phosphorous, respectively, thus indicating the ordered arrangement of O and F atoms below RT.

\begin{table}
\caption{\label{tab:alpha}
Atomic positions, isotropic atomic displacement parameters ($U_{\iso}$, in 10$^{-2}$~\r A$^2$), and occupancy factors ($f$) of $\alpha$-\navpof\ at 560~K (upper lines) and \mbox{$\gamma'$-\navpof} at 150~K (bottom lines). The $f$ values are fractions of total occupancy for a given position.
}
\begin{ruledtabular}
\begin{tabular}{ccccccc}
  Atom & Position & $x$          & $y$          & $z$        & $U_{\iso}$ & $f$     \\
  Na1  & $8h$     & 0.2651(5)    & 0.2651(5)    & 0          & 2.9(2)     & 0.52(1) \\
       &          & 0.2627(8)    & 0.2627(8)    & 0          & 0.6(1)     & 0.49(2) \\
  Na2  & $16l$    & 0.394(3)     & 0.209(2)     & 0          & 3.5(5)     & 0.11(1) \\ 
       &          & 0.336(3)     & 0.229(2)      & 0          & 0.6(1)     & 0.13(1) \\
  V    & $4e$     & 0            & 0            & 0.1982(1)  & 0.94(3)    & 1       \\
       &          & 0            & 0            & 0.1996(1)  & 0.37(4)    & 1       \\
  P    & $4d$     & $\frac12$    & 0            & $\frac14$  & 0.78(4)    & 1       \\
       &          & $\frac12$    & 0            & $\frac14$  & 0.02(6)    & 1       \\
  O1   & $4e$     & 0            & 0            & 0.3507(3)  & 1.8(1)     & 1       \\
       &          & 0            & 0            & 0.3555(4)  & 0.22(6)    & 1       \\
  O2   & $16n$    & 0.3091(2)    & 0            & 0.1634(1)  & 0.98(5)    & 1       \\
       &          & 0.3082(4)    & 0            & 0.1626(2)  & 0.22(6)    & 1       \\
  F    & $2a$     & 0            & 0            & 0          & 1.2(1)     & 1       \\
       &          & 0            & 0            & 0          & 0.22(6)    & 1       \\
\end{tabular}
\end{ruledtabular}
\end{table}
\begin{table*}
\begin{minipage}{14cm}
\caption{\label{tab:distances}
Interatomic distances (in~\r A) in $\alpha$-, $\beta$-, and $\gamma$-polymorphs of \navpof\ at 560~K, 298~K, and 150~K, respectively.
}
\begin{ruledtabular}
\begin{tabular}{crcrcr}
  \multicolumn{2}{c}{$\alpha$} & \multicolumn{2}{c}{$\beta$} & \multicolumn{2}{c}{$\gamma$} \\
  V--O1 & 1.626(3)           & V--O1  & 1.636(2)           & V--O1  & 1.629(3)           \\
  V--O2 & $4\times 2.011(1)$ & V--O2  & $2\times 2.001(5)$ & V--O2  & 2.032(9)           \\
        &                    & V--O3  & 1.991(5)           & V--O3  & 1.986(9)           \\
        &                    & V--O4  & 2.023(5)           & V--O4  & 1.992(9)           \\
        &                    &        &                    & V--O5  & 2.01(1)            \\
  P--O2 & $4\times 1.530(1)$ & P1--O2 & $4\times 1.543(6)$ & P1--O2 & $4\times 1.549(8)$ \\
        &                    & P2--O3 & $2\times 1.540(7)$ & P2--O3 & $4\times 1.542(8)$ \\
        &                    & P2--O4 & $2\times 1.509(7)$ & P3--O4 & $2\times 1.502(8)$ \\
        &                    &        &                    & P3--O5 & $2\times 1.550(9)$ \\
\end{tabular}
\end{ruledtabular}
\end{minipage}
\end{table*}

While the O and F atoms in \navpof\ form an ordered framework, the Na atoms are disordered. Our refinement for the $\alpha$-modification identifies two Na positions, $8h$ (Na1) and $16l$ (Na2). After constraining the sum of occupancies to the \navpof\ composition and refining atomic displacement parameters (ADPs) together with the occupancies, we found that the $8h$ position is approximately half-filled, whereas the $16l$ position is filled by $\frac18$ (i.e., twice less Na atoms than in $8h$).\cite{note2} This is in line with Ref.~\onlinecite{massa2002} that reported the occupancies of $\frac38$ and $\frac{3}{16}$ for $8h$ and $16l$, respectively (our experiment is done at 560~K compared to 300~K in Ref.~\onlinecite{massa2002}, hence a redistribution of Na atoms is possible). By contrast, Ref.~\onlinecite{sauvage2006} assigns Na2 to a $8j$ position with a negligible occupancy of $\frac{1}{16}$, i.e., 0.5~atoms per unit cell in $8j$ compared to approximately 2~atoms in $16l$ in our refinement. 

\subsection{Room-temperature $\beta$-polymorph}
\begin{figure}
\includegraphics{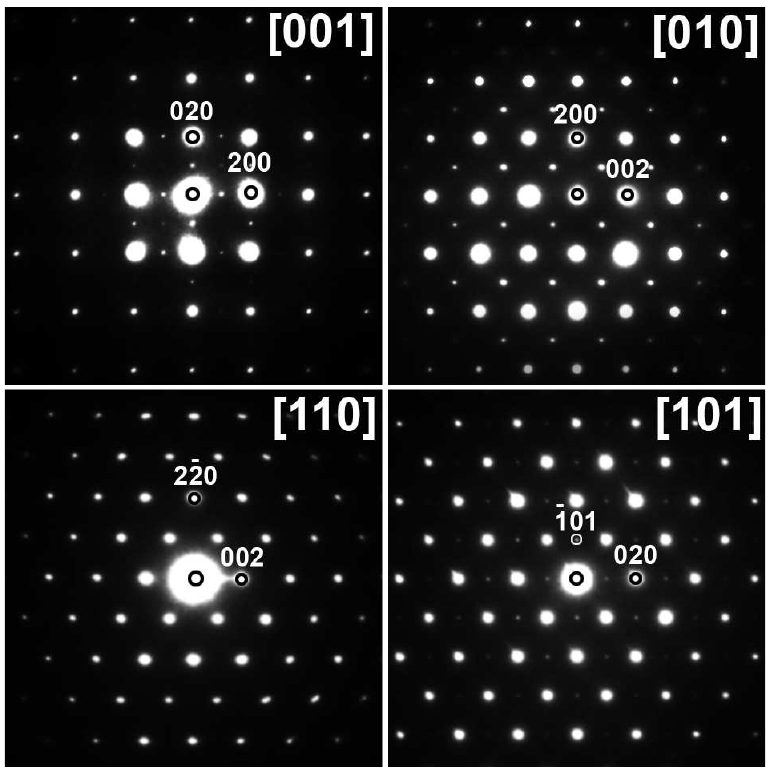}
\caption{\label{fig:ed-beta}
ED patterns of $\beta$-\navpof\ at RT. The patterns are indexed on a tetragonal $\sqrt2\,a_{\sub}\times\sqrt2\,a_{\sub}\times c$ supercell. Bright reflections correspond to the $a_{\sub}\times a_{\sub}\times c$ subcell, whereas weak reflections indicate the supercell.
}
\end{figure}

The superstructure formation in the RT $\beta$-polymorph is confirmed by ED (Fig.~\ref{fig:ed-beta}). While the intense reflections are assigned to the body-centered tetragonal \mbox{$a_{\sub}\times a_{\sub}\times c$} unit cell of the $\alpha$-polymorph, weaker superstructure reflections are clearly visible in the [001], [010], and [101] patterns. XRD and ED suggest a primitive tetragonal $\sqrt2\,a_{\sub}\times\sqrt2\,a_{\sub}\times c$ unit cell, with the reflection condition $h0l$, $h+l=2n$ that identifies the $P4_2/mnm$ space group. This reflection condition is present in the [010] pattern (Fig.~\ref{fig:ed-beta}). The emergence of the forbidden reflections, such as 100 and 010 in the [001] pattern, is caused by the multiple diffraction. Finally, the [110] pattern reveals sharper reflection conditions $h+k=2n$ and $h+l=2n$ corresponding to the body-centered unit cell of the $\alpha$-polymorph. The lack of superstructure reflections, such as $1\bar1 0$, in the [110] ED pattern is due to their zero structure factors, in agreement with the complete absence of these reflections in the XRD pattern. 

The superstructure formation at room temperature is related to the partial ordering of the Na atoms in the $\beta$-modification. The structure refinement (Table~\ref{tab:beta})\cite{supplement} showed that the VOPO$_4$F$_{0.5}$ framework remains intact (Table~\ref{tab:distances}), whereas Na atoms are partially ordered in two $8i$ positions. The Na1 position is completely filled, and the Na2 position is exactly half-occupied under the constraint of the \navpof\ composition. 

\begin{figure}[!ht]
\includegraphics{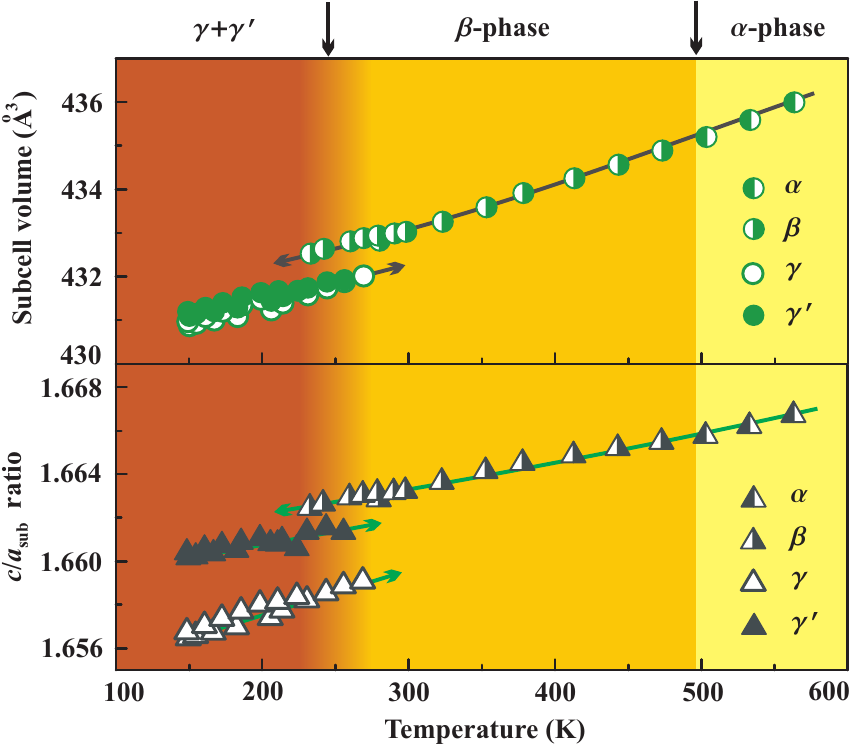}
\caption{\label{fig:cell}
(Color online) Temperature dependence of the subcell volume (top) and $c/a_{\sub}$ ratio (bottom) for \navpof. Error bars are smaller than symbols. Lines are guide for the eyes, with arrowheads showing changes upon heating/cooling in the region of the $\beta\leftrightarrow\gamma+\gamma'$ transition.\cite{note7}
}
\end{figure}

\begin{table}
\caption{\label{tab:beta}
Atomic positions and isotropic atomic displacement parameters ($U_{\iso}$, in 10$^{-2}$~\r A$^2$) of $\beta$-\navpof\ at 298~K. The Na2 position is half-filled.
}
\begin{ruledtabular}
\begin{tabular}{cccccc}
  Atom & Position & $x$          & $y$          & $z$        & $U_{\iso}$  \\
  Na1  & $8i$     & 0.5144(5)    & 0.2314(7)    & 0          & 2.8(1)      \\
  Na2  & $8i$     & 0.785(1)     & 0.028(1)     & 0          & 6.0(4)      \\
  V    & $8j$     & 0.2484(1)    & 0.2484(1)    & 0.80118(6) & 0.60(1)     \\
  P1   & $4d$     & 0            & $\frac12$    & $\frac14$  & 0.6(1)      \\
  P2   & $4e$     & 0            & 0            & 0.2558(3)  & 0.2(1)      \\
  O1   & $8j$     & 0.2499(4)    & 0.2499(4)    & 0.3529(2)  & 1.3(1)      \\
  O2   & $16k$    & 0.0957(6)    & 0.4031(6)    & 0.1623(6)  & 0.7(2)      \\
  O3   & $8j$     & 0.0949(6)    & 0.0949(6)    & 0.1662(7)  & 0.5(2)      \\
  O4   & $8j$     & 0.4037(6)    & 0.4037(6)    & 0.1617(8)  & 1.0(3)      \\
  F    & $4f$     & 0.2450(4)    & 0.2450(4)    & 0          & 0.8(1)      \\
\end{tabular}
\end{ruledtabular}
\end{table}

The structure refinements for $\alpha$- and $\beta$-\navpof\ suggest that the phase transition at 500~K is of the second order (order-disorder type). Indeed, the temperature evolution of lattice parameters (Fig.~\ref{fig:cell}) shows a smooth change in the cell volume around 500~K. The structural change also conforms to symmetry requirements for a second-order transition: the $P4_2/mnm$ space group can be derived from $I4/mmm$ using the $X_3^-$ irreducible representation. Similar to second-order transitions in (CuCl)LaNb$_2$O$_7$ (Ref.~\onlinecite{cucl}), we did not observe the $\alpha\leftrightarrow\beta$ transformation by DSC, presumably, due to the small change in the entropy.
\begin{figure}
\includegraphics{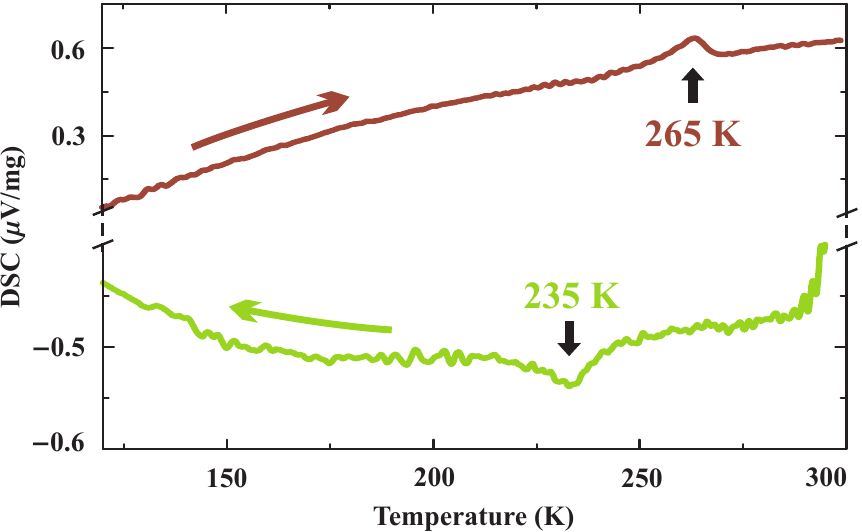}
\caption{\label{fig:dsc}
(Color online) DSC data showing the first-order \mbox{$\beta\leftrightarrow\gamma+\gamma'$} transition at 265~K (heating, upper panel) and 235~K (cooling, bottom panel).
}
\end{figure}

\subsection{Low-temperature phase separation}
Below RT, one can expect a further ordering of the Na atoms. Indeed, we found another structural transformation around 250~K. This transition is of the first-order type, as shown by: i) the temperature hysteresis (about 235~K on cooling and about 265~K on heating, Fig.~\ref{fig:dsc}); ii) the coexistence of the low-temperature and high-temperature phases in a certain temperature range; iii) the abrupt change in the cell volume (Fig.~\ref{fig:cell}). The transition is also revealed by an abrupt change in the ESR linewidth (see Sec.~\ref{sec:esr}).

\begin{table}
\caption{\label{tab:gamma}
Atomic positions and isotropic atomic displacement parameters ($U_{\iso}$, in 10$^{-2}$~\r A$^2$) of $\gamma$-\navpof\ at 150~K. 
}
\begin{ruledtabular}
\begin{tabular}{cccccc}
  Atom & Position & $x$          & $y$          & $z$        & $U_{\iso}$  \\
  Na1  & $8h$     & 0.3715(4)    & 0.6291(4)    & 0          & 0.69(5)     \\
  Na2  & $8h$     & 0.1186(4)    & 0.5951(4)    & 0          & 0.69(5)     \\
  Na3  & $8h$     & 0.3486(4)    & 0.3764(3)    & 0          & 0.69(5)     \\
  V    & $16i$    & $-0.0021(1)$ & 0.2511(1)    & 0.1991(1)  & 0.35(3)     \\
  P1   & $4b$     & 0            & 0            & $\frac14$  & 0.29(3)     \\
  P2   & $4d$     & 0            & $\frac12$    & $\frac14$  & 0.29(3)     \\
  P3   & $8g$     & 0.2490(2)    & 0.7490(2)    & $\frac14$  & 0.29(3)     \\
  O1   & $16i$    & $-0.0144(3)$ & 0.2440(4)    & 0.3522(3)  & 0.28(4)     \\
  O2   & $16i$    & 0.0957(7)    & 0.0079(4)    & 0.3400(6)  & 0.28(4)     \\
  O3   & $16i$    & 0.5965(7)    & $-0.0044(4)$ & 0.3379(6)  & 0.28(4)     \\
  O4   & $16i$    & 0.2583(3)    & 0.3476(7)    & 0.3261(6)  & 0.28(4)     \\
  O5   & $16i$    & 0.7567(4)    & 0.3440(8)    & 0.3436(6)  & 0.28(4)     \\
  F    & $8h$     & $-0.0045(4)$ & 0.2463(4)    & 0          & 0.28(4)     \\
\end{tabular}
\end{ruledtabular}
\end{table}

\begin{figure}
\includegraphics{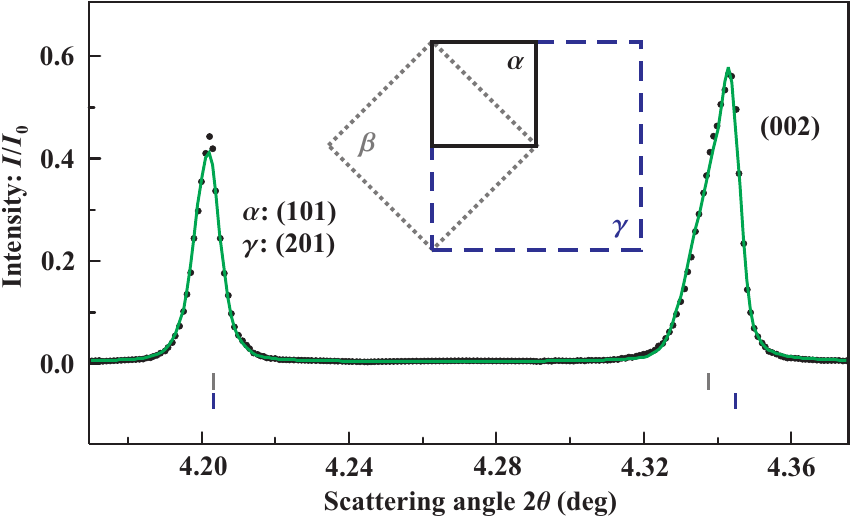}
\caption{\label{fig:002}
(Color online) XRD pattern of \navpof\ collected at 150~K (black dots) and the two-phase refinement (solid line). Upper and lower ticks denote the reflections of the $\gamma'$- and $\gamma$-phases. The inset shows the unit cells of $\alpha$- (solid), $\beta$- (dotted), and $\gamma$- (dashed) polymorphs.
}
\end{figure}

Upon cooling the sample, new reflections appeared below 235~K, and the reflections of the $\beta$-phase fully disappeared at 210~K. Below 210~K, the pattern could be indexed on a $2a_{\sub}\times 2a_{\sub}\times c$ tetragonal unit cell that we further refer as $\gamma$-modification. However, a closer examination showed a complex shape of the $(hkl)$ reflections with non-zero $l$. In particular, the $(002)$ reflection at $2\theta=4.34^{\circ}$ could not be fitted as a single peak (Fig.~\ref{fig:002}). This points to the coexistence of two phases with slightly different lattice parameters.

\begin{figure}
\includegraphics{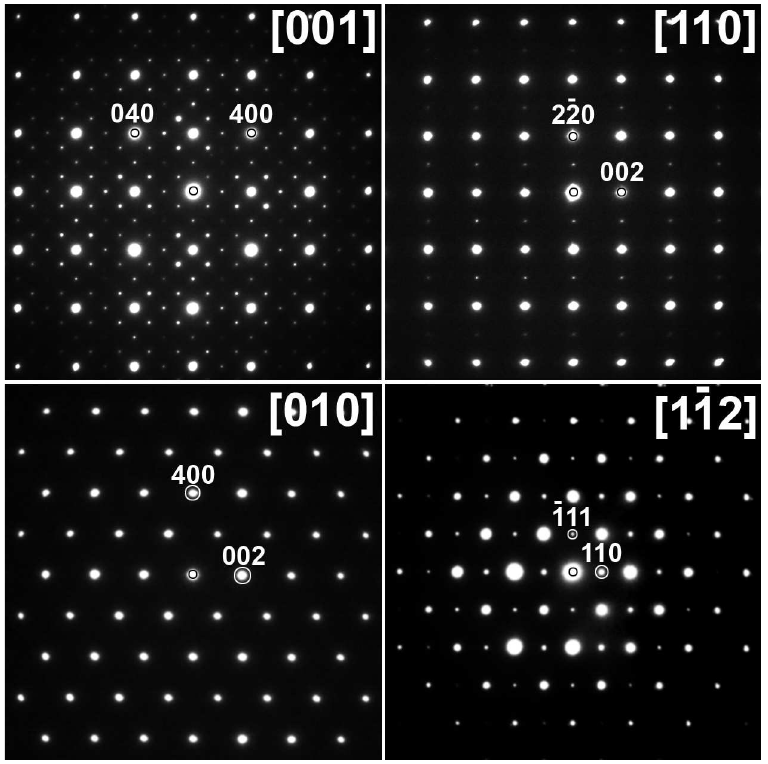}
\caption{\label{fig:ed-gamma1}
ED patterns of $\gamma$-\navpof\ at 100~K. The patterns are indexed on a tetragonal $2a_{\sub}\times 2a_{\sub}\times c$ unit cell. Brighter reflections with even $h$, $k$, and $l$ indices arise from the subcell, whereas weaker spots evidence the formation of the superstructure.
}
\end{figure}

\begin{figure}
\includegraphics{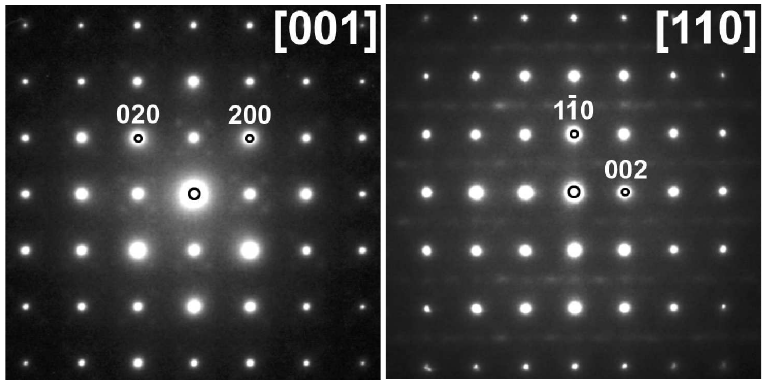}
\caption{\label{fig:ed-gamma2}
ED patterns of $\gamma'$-\navpof\ at 100~K. The patterns are indexed on a body-centered tetragonal \mbox{$a_{\sub}\times a_{\sub}\times c$} unit cell. Diffuse lines in the [110] pattern evidence short-range order.
}
\end{figure}
The low-temperature ED study confirmed the presence of two different phases below RT. One of the phases is the ordered $\gamma$-polymorph that shows sharp superstructure reflections of the $2a_{\sub}\times 2a_{\sub}\times c$ unit cell (Fig.~\ref{fig:ed-gamma1}). The crystallites of the second phase are less ordered and reveal diffuse satellites or diffuse intensity lines instead of the superstructure reflections (Fig.~\ref{fig:ed-gamma2}). The $\gamma$-polymorph has a primitive tetragonal unit cell. The reflection conditions $hhl$, $l=2n$ ([110] pattern) and $h0l$, $h=2n$ ([001] and [010] patterns) were confirmed by XRD, and resulted in the $P4_2/mbc$ space group. Similar to the [110] pattern of $\beta$-\navpof\ (Fig.~\ref{fig:ed-beta}), the [010] pattern of the $\gamma$-polymorph contains subcell reflections only. 

The second low-temperature phase, further referred as $\gamma'$-polymorph, reveals sharp subcell reflections as well as diffuse satellites on the [001] pattern and diffuse lines along $c^*$ on the [110] pattern (Fig.~\ref{fig:ed-gamma2}). The intensity of the diffuse lines is modulated in a way that the maxima coincide with the superstructure reflections of the $\gamma$-phase. This suggests that the two low-temperatures polymorphs develop a similar supercell, with the long-range order in $\gamma$ and the short-range order in $\gamma'$. While the ordered structure of $\gamma$-\navpof\ can be determined from XRD (see below), the nature of the short-range order in $\gamma'$-\navpof\ is difficult to establish because the compound is unstable under the electron beam.

In XRD, the diffuse intensity produced by the \mbox{$\gamma'$-polymorph} is smeared. The remaining subcell reflections are formally equivalent to that of the $\alpha$-phase. The 150~K pattern could be refined as a two-phase mixture of the $\gamma$-phase with the $2a_{\sub}\times 2a_{\sub}\times c$ unit cell and the $\gamma'$-phase with the $a_{\sub}\times a_{\sub}\times c$ unit cell. The refined composition of the mixture yields about 40~\% of the $\gamma'$-phase. This ratio is nearly constant within the temperature range under investigation.

The structure refinement of $\gamma$-\navpof\ (Table~\ref{tab:gamma})\cite{supplement} identifies three Na positions which are fully occupied, as evidenced by their low ADPs.\cite{note3} The ordering of Na has moderate effect on the framework: the V--O and P--O distances are nearly unchanged compared to the $\alpha$- and $\beta$-phases (Table~\ref{tab:distances}). The refined structure of $\gamma'$-\navpof\ closely resembles the $\alpha$-polymorph (Table~\ref{tab:alpha}). Based on the XRD and ED data, we suggest that \navpof\ develops a two-phase mixture at the first-order transition around 250~K. Specific heat, DSC, and ESR measurements (Sec.~\ref{sec:esr} and Ref.~\onlinecite{supplement}) do not reveal any further structural changes below this temperature.

At low temperatures, the $\gamma$- and $\gamma'$-phases have the same subcell volume but a different $c/a_{\sub}$ ratio (Fig.~\ref{fig:cell}). Based on our synchrotron XRD data as well as magnetization, NMR, and ESR measurements (Sec.~\ref{sec:experiment}), we conjecture that both phases contain V$^{+4}$, and their composition is, therefore, similar (within the available resolution). The origin of the phase separation in \navpof\ is presently unclear and should be disclosed in future studies. For a further discussion on Na ordering in \navpof\ and related compounds, we refer the reader to Sec.~\ref{sec:disc-cryst}.
\section{Magnetic behavior}
\label{sec:experiment}
\subsection{Magnetization}
\label{sec:mag}
\begin{figure}
\includegraphics{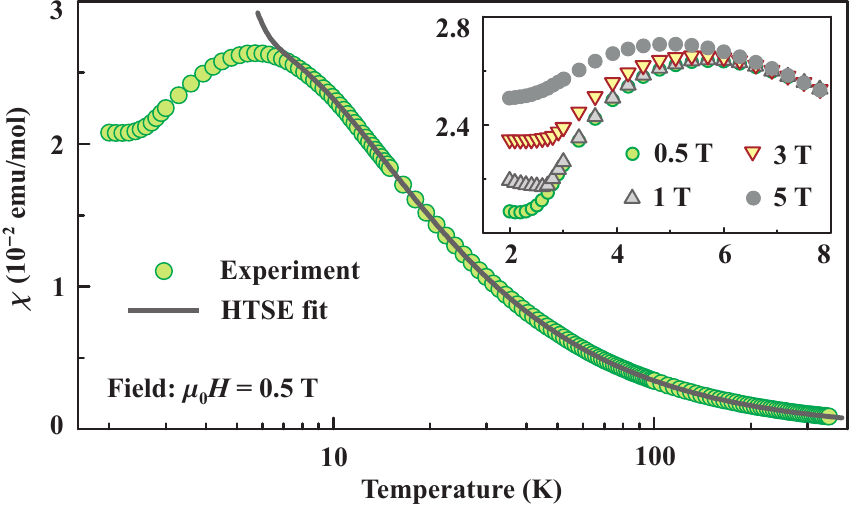}
\caption{\label{fig:magn}
(Color online) Temperature dependence of the magnetic susceptibility ($\chi$) and the HTSE fit with Eq.~\eqref{eq:htse}: $T_{\min}=12$~K, $\bar J_1=-3.7(1)$~K, $\bar J_2=6.6(1)$~K. The inset shows the field dependence of $\chi$ at low temperatures. The kink at $2.6-2.8$~K denotes the onset of the long-range order, see also Fig.~\ref{fig:diagram}.
}
\end{figure}

The temperature dependence of the magnetic susceptibility ($\chi$) of \navpof\ is typical for a low-dimensional (and possibly frustrated) antiferromagnet, see Fig.~\ref{fig:magn}. At low fields ($\mu_0H\leq 1$~T), the maximum at $T^{\chi}_{\max}\simeq 5.7$~K manifests the onset of AFM short-range correlations followed by a kink at $T_N\simeq 2.6$~K due to the long-range ordering (see also the specific heat and NMR spin-lattice relaxation rate data in Figs.~\ref{fig:heat} and~\ref{fig:t1}, respectively). A stronger magnetic field has a pronounced effect on $\chi$: the maximum shifts to lower temperatures, whereas $T_N$ slightly increases (see also Fig.~\ref{fig:diagram}). Above 50~K, $1/\chi(T)$ shows a nearly linear behavior indicating the Curie-Weiss paramagnetic regime. No signatures of the $\beta\leftrightarrow\gamma+\gamma'$ transition around 250~K could be observed.

Above 50~K, the fit of the data with the Curie-Weiss law:
\begin{equation}
  \chi=\chi_0+\dfrac{C}{T+\theta}
\label{eq:cw}\end{equation}
yields temperature-independent susceptibility $\chi_0=-1.5(1)\times 10^{-4}$~emu/mol (core diamagnetism and Van Vleck paramagnetism), the Curie constant $C=0.366(2)$~emu~K/mol, and the Curie-Weiss temperature $\theta=3.3(1)$~K. The effective magnetic moment of 1.71(1)~$\mu_B$ conforms to the expected value of $g\mu_B\sqrt{S(S+1)}=1.69$~$\mu_B$ for V$^{+4}$ where we used $S=\frac12$ and the powder-averaged $g=1.95$ from ESR (see Sec.~\ref{sec:esr}).

Despite the 3D nature of the crystal structure, a low-dimensional magnetic behavior should be expected. Previous studies of V$^{+4}$ compounds\cite{nath2008b,tsirlin2010,mazurenko2006} identify the typical $d_{xy}$ orbital ground state driven by the short V--O1 bond along the $z$ direction. Since the $d_{xy}$ orbital does not overlap with the orbitals of the axial O1 and F atoms, exchange couplings along the $c$ direction are weak. In the case of \navpof, this implies the formation of VOPO$_4$ magnetic layers, similar to $AA'$VO(PO$_4)_2$ FSL-type vanadium phosphates (see the left panel of Fig.~\ref{fig:structure}).\cite{kaul2004,nath2008,tsirlin2009} Three different phosphorus positions in the $\gamma$-\navpof\ structure lead to a number of inequivalent exchange couplings (Sec.~\ref{sec:band} and Fig.~\ref{fig:gamma}). However, the thermodynamic properties in the high-temperature region ($T>|J_1|,J_2$) should follow the predictions for the ideal FSL model, and allow one to evaluate $\bar J_1$ and $\bar J_2$, the averaged nearest-neighbor (NN) and next-nearest-neighbor (NNN) couplings on the square lattice.\cite{tsirlin2009}

\begin{figure}
\includegraphics{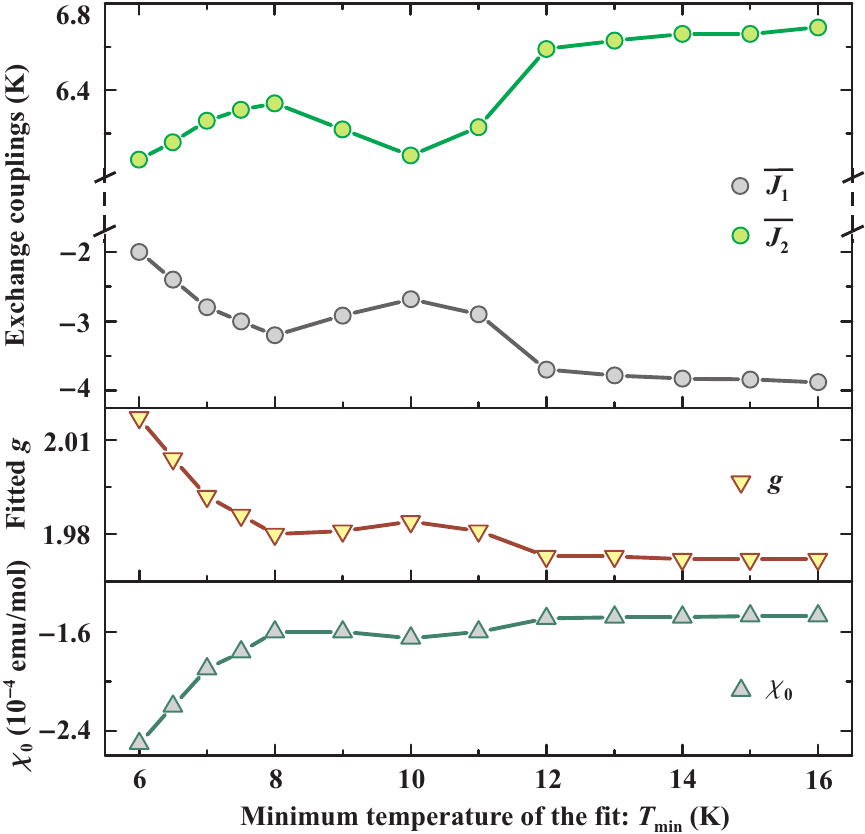}
\caption{\label{fig:fits}
(Color online) Details of the magnetic susceptibility fit: averaged exchange couplings $\bar J_1$ and $\bar J_2$, $g$-value, and temperature-independent contribution $\chi_0$ evaluated for the fitting ranges with different $T_{\min}$.
}
\end{figure}

Based on the above arguments, we fit the magnetic susceptibility of \navpof\ with the high-temperature series expansion (HTSE) for the FSL model:\cite{rosner2003}
\begin{equation}
  \chi=\chi_0+\dfrac{N_Ag^2\mu_B^2}{k_BT}\sum_n\left(\dfrac{\bar J_1}{k_BT}\right)^n\sum_m c_{mn}\left(\dfrac{\bar J_2}{\bar J_1}\right)^m,
\label{eq:htse}\end{equation}
where $\chi_0$ is similar to Eq.~\eqref{eq:cw}, $N_A$ is Avogadro's number, $g$ is the $g$-factor, and $c_{mn}$ with $m,n\leq 9$ are coefficients of the HTSE. Owing to the internal symmetry of the model, one typically finds two equivalent fits with different $\bar J_1$ and $\bar J_2$ (see Refs.~\onlinecite{kaul2004,kaul-thesis,nath2008,tsirlin2010}). For example, fitting the data above 12~K, we arrive at FM $\bar J_1=-3.7(1)$~K and AFM $\bar J_2=6.6(1)$~K ($g=1.973(2)$, $\chi_0=-1.5(1)\times 10^{-4}$~emu/mol) or AFM $\bar J_1=5.5(1)$~K and FM $\bar J_2=-2.1(1)$~K ($g=1.981(2)$, $\chi_0=-1.6(1)\times 10^{-4}$~emu/mol). The correct solution can be determined from the specific heat data,\cite{kaul-thesis} the saturation field,\cite{nath2008} or band structure calculations.\cite{tsirlin2009,tsirlin2010} In the case of \navpof, band structure calculations provide a solid evidence for leading AFM couplings between next-nearest neighbors (see Sec.~\ref{sec:band}). The analysis of the saturation field also favors the solution with FM $\bar J_1$ and AFM $\bar J_2$ (see below).

A more subtle problem of the HTSE fit lies in the choice of the fitting range. The convergence of the HTSE depends on the frustration ratio $\alpha$, hence the lower limit of the fitting range ($T_{\min}$) is not universal. Fitting the data with different $T_{\min}$, we find a stable solution for $T_{\min}=12-16$~K (Fig.~\ref{fig:fits}). The resulting $\chi_0=-1.5(1)\times 10^{-4}$~emu/mol and $g=1.973(2)$ are in excellent agreement with the Curie-Weiss fit ($\chi_0=-1.5(1)\times 10^{-4}$~emu/mol, $g=1.97(1)$). At lower $T_{\min}$, the high-temperature part of the HTSE fit deviates from the Curie-Weiss fit, thereby less accurate estimates of $\bar J_1$ and $\bar J_2$ are obtained. Using $\bar J_1\simeq -3.7$~K and $\bar J_2\simeq 6.6$~K,
we arrive at $\alpha=\bar J_2/\bar J_1\simeq -1.8$ that coincides with the frustration regimes of Pb$_2$VO(PO$_4)_2$, BaZnVO(PO$_4)_2$, and PbZnVO(PO$_4)_2$, see Table~\ref{tab:comparison} as well as Refs.~\onlinecite{nath2008,nath2009}.

The magnetization curve of \navpof\ (right panel of Fig.~\ref{fig:diagram})\cite{high-field} is typical for a two-dimensional (2D) antiferromagnet. The linear increase in the magnetization ($M$) at low fields is followed by a slight curvature and the saturation at $\mu_0H_s\simeq 15.4$~T. To test our estimates of $\bar J_1$ and $\bar J_2$ against $H_s$, we use the expression from Ref.~\onlinecite{schmidt2007} for the columnar AFM state ($J_2>-0.5J_1$) on the regular square lattice:
\begin{equation}
  \mu_0H_s=2(\bar J_1+2\bar J_2)k_B/(g\mu_B).
\label{eq:hs}\end{equation}
Assuming $g=1.95$ from ESR (Sec.~\ref{sec:esr}), one finds $\mu_0H_s\simeq 14.5$~T which is about 1~T lower than the experimental value. The solution with AFM $J_1$ implies the N\'eel ordering (antiparallel spins on nearest neighbors), leads to $\mu_0H_s\simeq 16.8$~T, and overestimates $H_s$, which in this case is proportional to $4\bar J_1$ instead of $2(\bar J_1+2\bar J_2)$ in Eq.~\eqref{eq:hs}.\cite{schmidt2007} While the saturation field itself does not allow us to choose the correct solution, DFT results (Sec.~\ref{sec:band}) put forward the sizable AFM $\bar J_2$ and, therefore, FM $\bar J_1$ -- AFM $\bar J_2$ regime with an additional interlayer coupling. This interlayer coupling contributes to the saturation field, and explains the underestimate of $H_s$ in the purely 2D model described by Eq.~\eqref{eq:hs}. By contrast, the AFM $\bar J_1$ -- FM $\bar J_2$ regime overestimates $H_s$ even in the 2D model, thereby any additional couplings will only exaggerate the discrepancy between the exchange integrals and saturation field. This way, we choose the FM $\bar J_1$ -- AFM $\bar J_2$ solution, which is confirmed by the similarity to other FSL-type vanadium phosphates.\cite{kaul-thesis,nath2008,tsirlin2010}
\subsection{Heat capacity}
\begin{figure}
\includegraphics{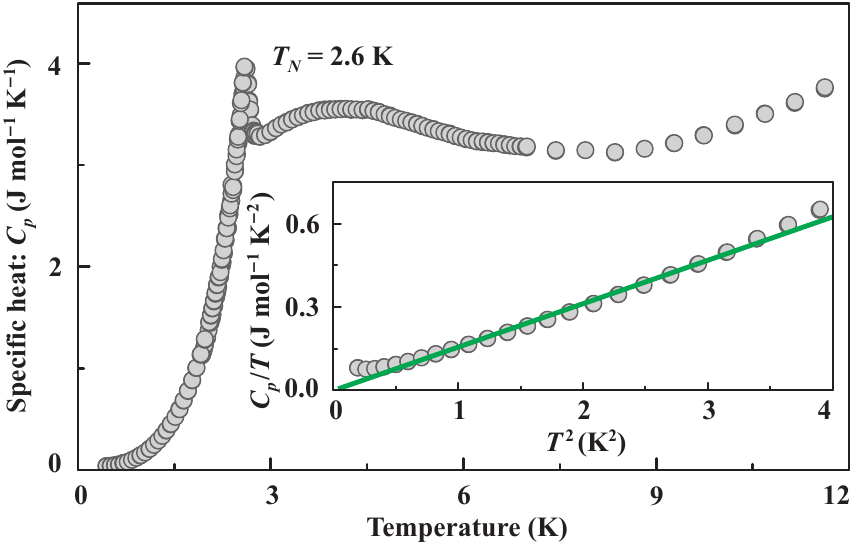}
\caption{\label{fig:heat}
(Color online) Zero-field specific heat ($C_p$) of \navpof\ with a transition anomaly at $T_N=2.6$~K and the maximum of the magnetic contribution around 4~K. The inset shows the low-temperature data plotted as $C_p/T$ against $T^2$. Solid line indicates the idealized $T^3$ behavior.
}
\end{figure}
The measured specific heat ($C_p$)\cite{supplement} of \navpof\ resembles that of BaCdVO(PO$_4)_2$ and other FSL compounds.\cite{kaul-thesis,nath2008,tsirlin2010} The sharp anomaly at $T_N\simeq 2.6$~K is followed by a maximum around 4~K and an increase in $C_p$ above 8~K (Fig.~\ref{fig:heat}). The maximum is a signature of the magnetic contribution $C_{\mg}$, whereas the data above 8~K are dominated by the phonon contribution. Unfortunately, we were unable to fit the high-temperature part with Debye functions and extract the magnetic contribution. The fitting problem may be related to the coexistence of $\gamma$- and $\gamma'$-polymorphs having different phonon spectra. 

\begin{figure*}
\includegraphics{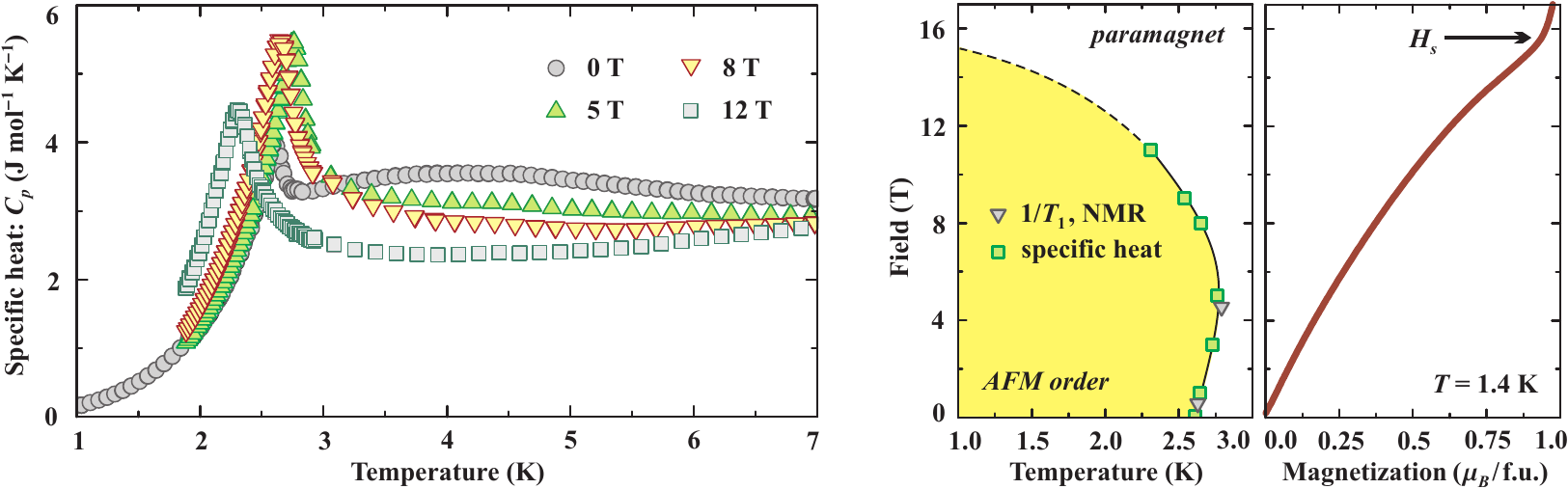}
\caption{\label{fig:diagram}
(Color online) Left: specific heat of \navpof\ measured in different applied fields. Right: the field-vs-temperature phase diagram, based on the heat capacity and NMR measurements, as well as the magnetization isotherm at 1.4~K (Ref.~\onlinecite{high-field}).
}
\end{figure*}
The specific heat maximum around 4.0~K suggests weaker exchange couplings in \navpof\ compared to Pb$_2$VO(PO$_4)_2$ and PbZnVO(PO$_4)_2$ that reveal $\bar J_2=9-10$~K and the maxima of $C_{\mg}$ at a higher temperature of $4.5-5.0$~K.\cite{kaul-thesis,tsirlin2010} By contrast, Li$_2$VOSiO$_4$ with a similar $J_2=6.3$~K has the maximum of $C_{\mg}$ at 3.6~K.\cite{melzi2001,kaul-thesis} Since the phonon contribution at 4~K is small, the measured value of $C_p$ is a reasonable estimate for the maximum of $C_{\mg}$. This way, we find $C_{\mg}^{\max}/R\simeq 0.43$ that compares well to 0.44, as expected for the unfrustrated\cite{sengupta2003} or moderately frustrated\cite{shannon2004} square lattice, and experimentally observed in a number of FSL compounds.\cite{kaul-thesis,tsirlin2010}

The low-temperature evolution of $C_p$ resembles classical antiferromagnets. However, the data do not follow a simple $T^3$ behavior (see the inset of Fig.~\ref{fig:heat}). The origin of this deviation is presently unclear. Note that other FSL-type compounds also show complex low-temperature features in the specific heat. The data for Pb$_2$VO(PO$_4)_2$ and Li$_2$VOGeO$_4$ can be described by a combination of cubic and linear terms.\cite{kaul-thesis} In \navpof, the linear term is vanishingly small, but the cubic term itself poorly fits the data below 2~K. 

Heat capacity measurements in magnetic field establish the temperature-vs-field phase diagram of \navpof\ (Fig.~\ref{fig:diagram}). With increasing magnetic field, the specific heat maximum is suppressed, whereas the transition anomaly is growing and $T_N$ increases up to 2.8~K at 5~T. Above 5~T, both the anomaly and $T_N$ are gradually suppressed. The overall shape of the phase boundary is typical for 2D antiferromagnets.\cite{nath2008,tsirlin2010,sengupta2009} The initial increase in $T_N$ is due to the field-induced anisotropy that weakens quantum fluctuations. Upon further increase in the field, the tendency towards the fully polarized state competes with AFM correlations and impedes the AFM long-range ordering. 
\subsection{Electron spin resonance}
\label{sec:esr}
\begin{figure}
\includegraphics[width=8cm]{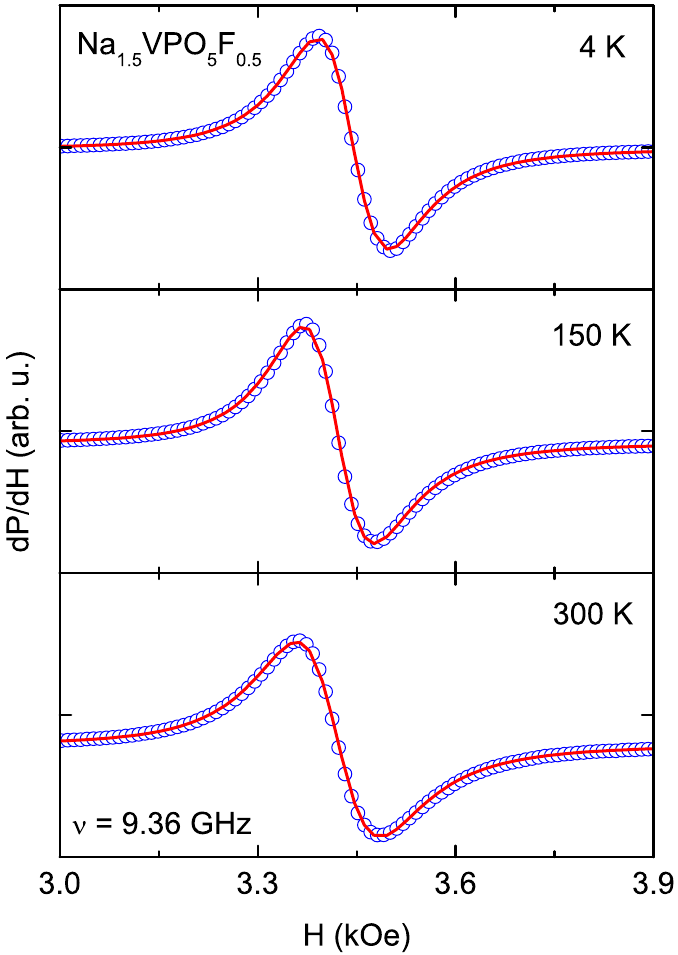}
\caption{\label{fig:spectra}
(Color online) ESR spectra of \navpof\ measured at X-band frequency for selected temperatures in the paramagnetic regime. Solid lines indicate fits with the field derivative of a Lorentzian.} 
\end{figure}
\begin{figure}
\includegraphics[width=8cm]{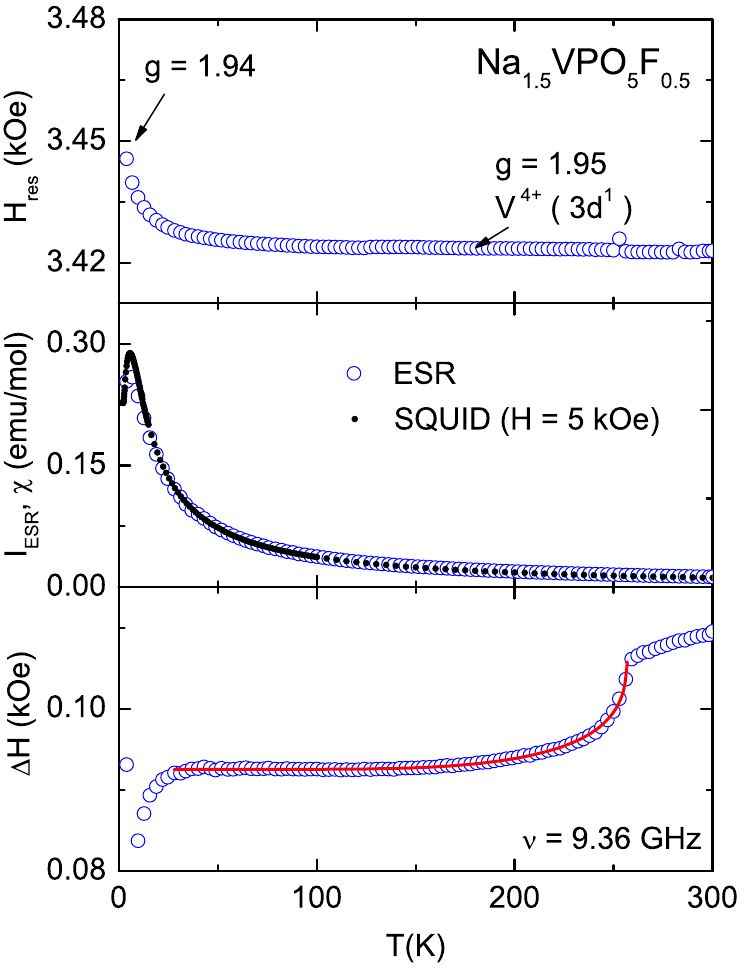}
\caption{\label{fig:esr}
(Color online) Temperature dependence of the ESR
parameters: resonance field $H_{\rm res}$ (upper frame), ESR intensity compared to
the magnetic susceptibility (middle frame), and linewidth $\Delta H$ (lower
frame) in \navpof. The solid red line indicates the fit using a mean-field BCS-like ansatz [Eq.~\eqref{eq:BCS}].} 
\end{figure}

Typical ESR spectra of \navpof\ are shown in Fig.~\ref{fig:spectra}. The compound reveals a single exchange-narrowed resonance line which is perfectly described by a Lorentzian curve in
the whole temperature range under investigation. The temperature dependence of the corresponding fit parameters, i.e., the resonance field $H_{\rm res}$, intensity $I_{\rm ESR}$, and half-width at half-maximum $\Delta H$, is illustrated in Fig.~\ref{fig:esr}. At room temperature, the
$g$-factor, derived from the resonance field $H_{\rm res}=3422$~Oe via $h\nu=g\mu_{\rm B}H_{\rm res}$ is found to be $g=1.95$, which is a typical value for V$^{4+}$ ($3d^{1}$) ions coordinated by oxygen atoms in a square pyramid or an octahedron.\cite{Abragam1970} On approaching $T_N$, the
$g$-value decreases only slightly. The integrated ESR intensity nicely matches the static susceptibility $\chi$ obtained from SQUID measurements, and exhibits a maximum around $7$~K. This indicates that all vanadium spins are observed by ESR. 

The linewidth is $\Delta H\simeq 110$~Oe at room temperature, and exhibits two anomalies upon cooling. One anomaly is close to $T_N$, probably due to short-range order effects, while the other one matches the structural $\beta\leftrightarrow\gamma+\gamma'$ transition at $T_{\str}\simeq 250$~K. Similar to the structural phase transition in the quasi one-dimensional magnet CuSb$_2$O$_6$,\cite{heinrich2003} the change in the ESR linewidth for $T<T_{\text{str}}$ is well described in terms of a thermally activated behavior:
\begin{equation}
\Delta H (T) = \Delta H_0 + A \cdot \exp{(-\frac{\Delta(T)}{T})}
\label{eq:BCS}
\end{equation}
with the BCS-like mean-field gap $\Delta(T) = 1.74 \Delta_0 \cdot(1-T/T_{\str})^{0.5}$. We obtain $\Delta_0 = 560(13)$~K, $T_{\str}=257(1)$~K, $A = 13(1)$~Oe, and the residual linewidth $\Delta H_0 = 93(1)$~Oe. It is interesting to consider the relation between the energy gap $\Delta_0$ and the transition temperature $T_{\text{str}}$, which yields $2 \Delta / T_{\str}=4.36$. This is enhanced compared to the value of 3.52 predicted by mean-field theory for a second-order phase transition. Indeed, the transition at $T_{\str}$ is of the first order, as shown by the temperature evolution of unit cell parameters (Fig.~\ref{fig:cell}).

\subsection{Nuclear magnetic resonance}
\label{sec:nmr}
\begin{figure}
\includegraphics[width=8.5cm]{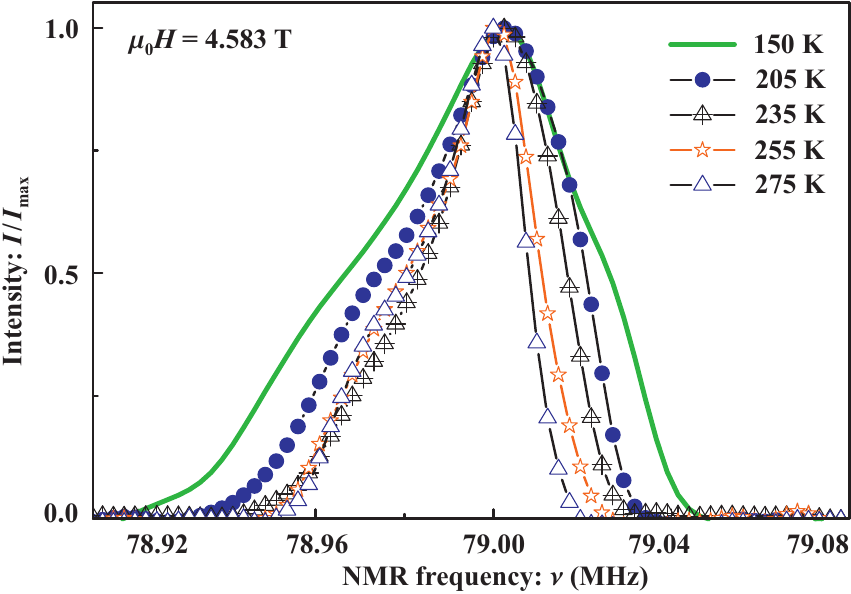}
\caption{\label{fig:nmr1}
(Color online) Fourier-transformed $^{31}$P NMR spectra at different temperatures $T$ ($T>T_N$) for \navpof.}
\end{figure}

For an $I=\frac12$ nucleus, one expects a single spectral line\cite{nath2005,nath2008b,nath2008c} that is indeed observed in the NMR spectra of \navpof\ in the $1.5-300$~K temperature range (Fig.~\ref{fig:nmr1}). Although our structural study (Sec.~\ref{sec:structure}) identifies two inequivalent phosphorous positions at 300~K ($\beta$-phase) and at least three inequivalent phosphorous positions in the $\gamma$-phase below $200-250$~K, this will not necessarily lead to separate lines in the spectra. The line shift for an individual P site is determined by its local environment and the hyperfine couplings to electronic spins on V$^{+4}$. Since the inequivalent P sites have very similar coordination and similar connections to the four surrounding vanadium atoms (Table~\ref{tab:distances} and Fig.~\ref{fig:gamma}), we believe that the line splitting is too small to be observed in the present experiment. While this does not allow to refine our model of the low-temperature crystal structure and phase separation, the single spectral line facilitates the study of the magnetic properties by NMR. 

\begin{figure}
\includegraphics{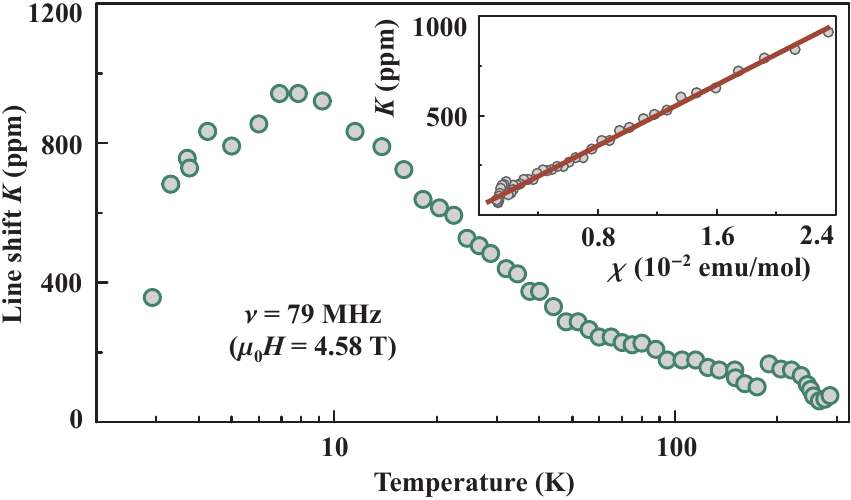}
\caption{\label{fig:nmr2}
(Color online) Temperature dependence of the $^{31}$P NMR line shift $K$ measured at 79~MHz. The inset shows $K$ plotted against the magnetic susceptibility $\chi$ with temperature as an implicit parameter. The solid line represents the linear fit with Eq.~\eqref{eq:shift}.}
\end{figure}
Since our measurements are done on randomly-oriented polycrystalline samples, the asymmetric shape of the spectra corresponds to a powder pattern due to an asymmetric hyperfine coupling constant and an anisotropic susceptibility. The line position was found to shift with temperature, similar to the magnetic susceptibility (Fig.~\ref{fig:nmr2}). To establish the relation between $K$ and $\chi$, we use the expression:\cite{nath2009}
\begin{equation}
K(T)=K_0+\frac{A_{\rm hf}}{N_A}\chi_{\rm spin}(T),
\label{eq:shift}
\end{equation}
where $K_0$ is the temperature-independent chemical shift, $A_{\rm hf}$ is the hyperfine coupling constant, and $\chi_{\rm spin}/N_A$ is the intrinsic (spin) magnetic susceptibility of \navpof\ in units of $\mu_B$/Oe per electronic spin (i.e., per formula unit). In order to calculate $A_{\rm hf}$, we plotted $K$ vs. $\chi$ with $T$ as an implicit parameter (the inset of Fig.~\ref{fig:nmr2}).\cite{note6} The linear fit with Eq.~\eqref{eq:shift} over the whole temperature range yields $A_{\rm hf} = (212 \pm 20)$ Oe/$\mu_B$ that is comparable to hyperfine couplings found in other FSL-type vanadium phosphates.\cite{nath2009}

\begin{figure}
\includegraphics[width=8.5cm]{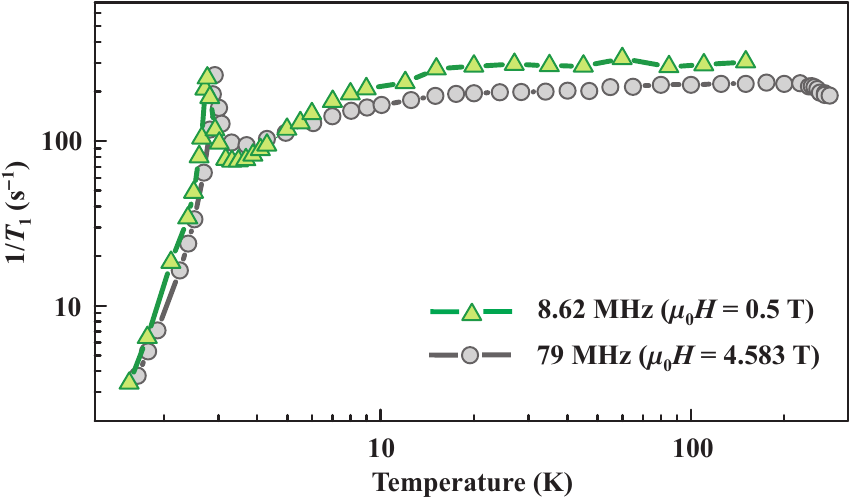}
\caption{\label{fig:t1} (Color online) Temperature dependence of the spin-lattice relaxation 
rate $1/T_{1}$ measured at two different frequencies of 79 and 8.62~MHz.
} 
\end{figure}
\begin{figure}
\includegraphics[width=8.5cm]{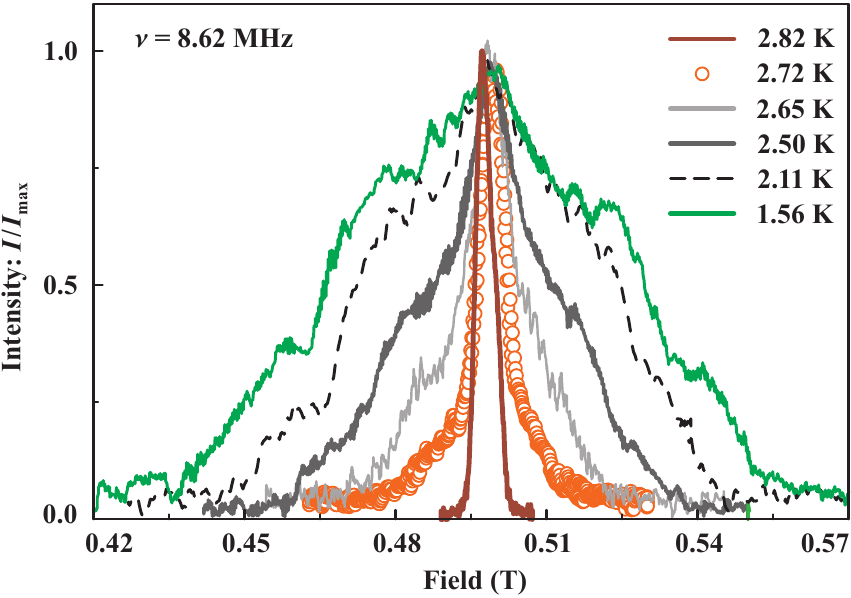}
\caption{\label{fig:belowTn} 
(Color online) Temperature-dependent $^{31}$P NMR spectra measured at 8.62~MHz.}
\end{figure}

For an $I=\frac12$ nucleus, the recovery of the longitudinal magnetization is expected to follow a single-exponential behavior. In \navpof, the recovery of the nuclear magnetization after a saturation pulse was indeed fitted well by the exponential function
\begin{equation}
1-\frac{M(t)}{M_0}=Ae^{-t/T_{1}},
\label{exp}
\end{equation}
where $M(t)$ is the nuclear magnetization at a time $t$ after the saturation pulse, and $M_0$ is the equilibrium magnetization. The temperature dependence of $1/T_1$ is depicted in Fig.~\ref{fig:t1}. Above 10~K, $1/T_1$ is essentially temperature-independent, which is typical for paramagnetic moments fluctuating fast and at random.\cite{moriya1956} Below 10~K, $1/T_1$ decreases slowly, shows a peak around 2.8~K, and further decreases at low temperatures. A similar behavior has been observed in Pb$_{2}$VO(PO$_4)_2$ from $^{31}$P NMR (Ref.~\onlinecite{nath2009}), SrZnVO(PO$_{4}$)$_{2}$ from $^{31}$P NMR (Ref.~\onlinecite{bossoni2010}), VOMoO$_4$ from $^{95}$Mo NMR (Ref.~\onlinecite{carretta2002}), and Cu(HCO$_2$)$_2\cdot 4$D$_2$O from $^{1}$H NMR (Ref.~\onlinecite{carretta2000}). The decrease in $1/T_1$ can be explained by the cancellation of antiferromagnetic spin fluctuations at the probed nuclei.\cite{carretta2000}


The sharp peak in $1/T_1$ identifies the onset of the long-range magnetic ordering. The peak position depends on the NMR frequency and reveals the field dependence of $T_N$. In Fig.~\ref{fig:diagram}, we plot the two transition temperatures derived from our NMR experiments, and find perfect agreement with the phase boundary obtained from the heat capacity measurements. 

Below $T_N$, the $^{31}$P line measured at 79~MHz broadens abruptly. In order to check whether any extra features could be resolved, we remeasured the spectra at a lower frequency of 8.62~MHz (Fig.~\ref{fig:belowTn}). Two symmetrical shoulder-like features develop on both sides of the central line and move systematically away from each other with decrease in temperature. The symmetric position and the systematic evolution of the shoulders in the low-frequency low-temperature spectra is an indication of a commensurate magnetic ordering. Our low-temperature spectral shape can be compared to the $^{7}$Li-NMR spectra of Li$_2$VOSiO$_4$ reported by Melzi \textit{et al.}\cite{melzi2000,melzi2001} In the columnar AFM phase, they observed a central line and two symmetrical satellites from the single-crystal measurement. In our case, such satellites appear as broad shoulders due to the random distribution of the internal field for the measurements on polycrystalline samples. Nevertheless, the overall spectral shape is consistent with $^{7}$Li-NMR results on Li$_{2}$VOSiO$_{4}$ and hence points to the columnar AFM ordering in \navpof.
\section{Microscopic magnetic model}
\label{sec:band}
\begin{figure}
\includegraphics{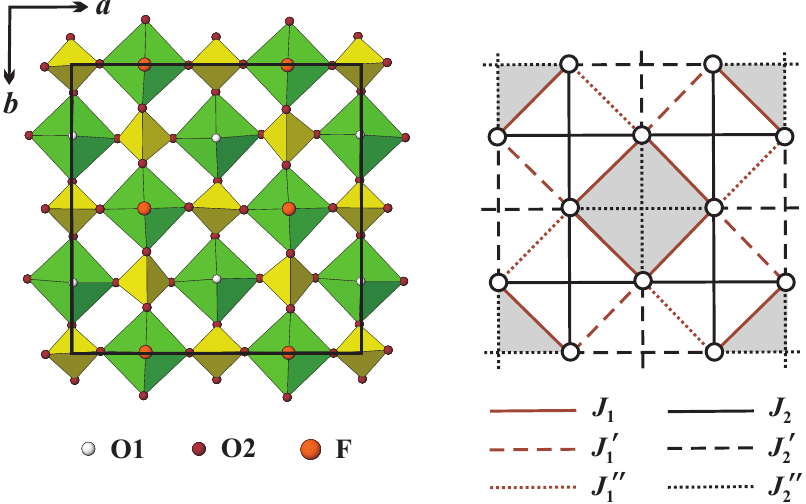}
\caption{\label{fig:gamma}
(Color online) Magnetic layers in the crystal structure of $\gamma$-\navpof\ (left) and the respective spin lattice with six inequivalent couplings (right). The $J_1-J_2''$ plaquette units are shaded.
}
\end{figure}

Despite the tetragonal crystallographic symmetry, the complex low-temperature superstructure of \mbox{$\gamma$-\navpof}\ gives rise to a number of inequivalent superexchange pathways and the overall distortion of the FSL. Using the atomic positions for $\gamma$-\navpof\ (Table~\ref{tab:gamma}), we arrive at six different couplings in the $ab$ plane: $J_1,J_1'$, and $J_1''$ between nearest neighbors as well as $J_2,J_2'$, and $J_2''$ between next-nearest neighbors, see Fig.~\ref{fig:gamma}. The spin lattice entails regular $J_1-J_2''$ plaquettes alternated with less regular plaquette-like units having one type of the NNN couplings (either $J_2$ or $J_2'$) yet different NN couplings. 

While it is hardly possible to evaluate all the six aforementioned parameters experimentally, DFT calculations are capable of providing microscopic insight into individual superexchange pathways and establishing realistic spin models for complex crystal structures.\cite{nath2008b,mazurenko2006,agvaso5} In the following, we present DFT results for the ordered structure of $\gamma$-\navpof. The second low-temperature phase, the $\gamma'$-polymorph with the short-range order of Na atoms (Sec.~\ref{sec:structure}), is more difficult to model due to the unresolved short-range order, although we may expect a similar scenario with FM couplings between nearest neighbors and AFM couplings between next-nearest neighbors.

\begin{figure}
\includegraphics{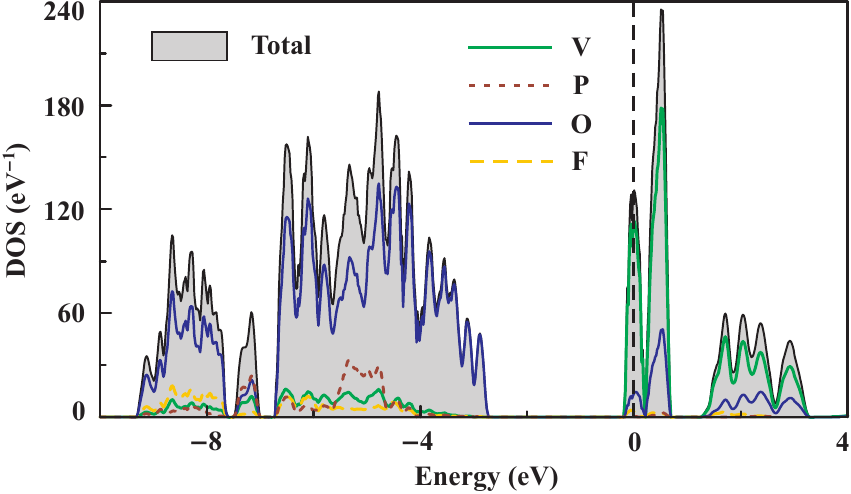}
\caption{\label{fig:dos}
(Color online) LDA density of states for $\gamma$-\navpof. The Fermi level is at zero energy.
}
\end{figure}

The LDA density of states (DOS) for \mbox{$\gamma$-\navpof} identifies oxygen and fluorine $2p$ valence bands below $-3$~eV and vanadium $3d$ bands at the Fermi level (Fig.~\ref{fig:dos}). The vanadium states show a crystal-field splitting into the $t_{2g}$ (below 0.7~eV) and $e_g$ (above 1.3~eV) sublevels, as expected for the octahedral coordination of a transition metal. Further on, the short bond to the axial oxygen atom (O1) splits the $t_{2g}$ bands into the half-filled $d_{xy}$ states lying at the Fermi level and unoccupied $d_{yz},d_{xz}$ states lying between 0.2~eV and 0.7~eV (here, $z$ aligns with the crystallographic $c$ axis). The positions of the crystal-field levels justify our empirical conclusions on the symmetry of the magnetic orbital that induces leading couplings in the $ab$ plane (Sec.~\ref{sec:mag}). The crystal-field splitting in $\gamma$-\navpof\ closely resembles the positions of the $3d$ sublevels in Pb$_2$V$_3$O$_9$ (Ref.~\onlinecite{pb2v3o9}), Ca(VO)$_2$(PO$_4)_2$ (Ref.~\onlinecite{nath2008b}), and other V$^{+4}$ oxides.

The gapless (metallic) nature of the energy spectrum is a typical problem of LDA due to the underestimate of correlation effects in the V $3d$ shell. While the correlation effects can be introduced in a mean-field way via the LSDA+$U$ method, the huge size of the unit cell (144 atoms) tempts us to avoid this procedure and apply an LDA-based model approach instead. The latter is remarkably efficient for the evaluation of weak exchange couplings in V$^{+4}$ compounds.\cite{mazurenko2006,tsirlin2010,pb2v3o9,agvaso5} Using Wannier functions with proper orbital characters,\cite{wannier} we fit 80 vanadium $3d$ bands with a tight-binding model, and map the resulting hopping parameters ($t$) onto the multi-orbital Hubbard model with the effective on-site Coulomb repulsion $U_{\eff}$ and on-site Hund's coupling $J_{\eff}$. The notably low $t$ values (below 5~meV) compared to $U_{\eff}\simeq 4$~eV identify the strongly correlated limit $t\ll U_{\eff}$ so that the expression of the Kugel-Khomskii model for the exchange couplings can be applied:\cite{kugel1982,mazurenko2006}
\begin{equation}
  J=\dfrac{4t_{xy}^2}{U_{\eff}}- \sum_{\alpha}\dfrac{4t_{xy\rightarrow\alpha}^2J_{\eff}}{(U_{\eff}+\Delta_{\alpha})(U_{\eff}+\Delta_{\alpha}-J_{\eff})}.
\label{eq:exchange}
\end{equation}
Here, $\alpha$ denotes unoccupied $d$ orbitals, $t_{xy}$ are hoppings between the $xy$ orbitals,  $t_{xy\rightarrow\alpha}$ are hoppings from the $xy$ (half-filled) to $\alpha$ (empty) orbitals, and $\Delta_{\alpha}$ are the crystal-field splittings between the $xy$ and $\alpha$ orbitals. The first term is AFM superexchange, whereas the second term is FM and arises from the Hund's coupling in the $3d$ shell.\cite{kugel1982,mazurenko2006,note1} Using $U_{\eff}=4$~eV and $J_{\eff}=1$~eV,\cite{pb2v3o9,agvaso5} we arrive at individual exchange couplings listed in Table~\ref{tab:exchange}.

The weak exchange couplings in \navpof\ are a severe challenge for computational methods. In contrast to PbZnVO(PO$_4)_2$ and Pb$_2$V$_3$O$_9$ (Refs.~\onlinecite{tsirlin2010,pb2v3o9}), the calculated $J$'s are exclusively AFM, despite the fits of the magnetic susceptibility (Sec.~\ref{sec:mag}) reveal the presence of FM exchange. This problem is likely unavoidable and restricts the DFT results to qualitative conclusions that are nevertheless essential for understanding the system. The sizable AFM NNN couplings, compared to weaker NN couplings, support the FM $\bar J_1$ -- AFM $\bar J_2$ scenario. The weakest NNN exchange $J_2''$ is about 70~\% of $J_2$ and $J_2'$ so that a spatial anisotropy of the diagonal couplings is moderate, as in Pb$_2$VO(PO$_4)_2$ and PbZnVO(PO$_4)_2$ (Refs.~\onlinecite{tsirlin2009,tsirlin2010}). By contrast, the difference between $J_1''$ and $J_1'$ ($J_1$) is about 5~K, which is larger than the experimental $|\bar J_1|\simeq 3.7$~K. This suggests a sizable spatial anisotropy of the NN couplings.

\begin{table}
\caption{\label{tab:exchange}
FM ($J^{\FM}$), AFM ($J^{\AFM}$), and total ($J$) exchange integrals calculated with Eq.~\eqref{eq:exchange}. The intralayer couplings are depicted in Fig.~\ref{fig:gamma}, whereas $J_{\perp}$ couples the neighboring layers (see Fig.~\ref{fig:structure}).
}
\begin{ruledtabular}
\begin{tabular}{ccccc}
          & Distance & $J^{\AFM}$ & $J^{\FM}$ & $J$  \\
          & (\r A)   & (K)        & (K)       & (K)  \\
  $J_1$   & 4.659    & 0.7        & $-0.1$    & 0.6  \\
  $J_1'$  & 4.586    & 1.2        & $-0.1$    & 1.1  \\
  $J_1''$ & 4.659    & 6.2        & 0         & 6.2  \\
  $J_2$   & 6.384    & 15.1       & 0         & 15.1 \\
  $J_2'$  & 6.356    & 13.5       & 0         & 13.5 \\
  $J_2''$ & 6.410    & 9.8        & 0         & 9.8  \\
  $J_{\perp}$ & 4.210 & 1.7       & 0         & 1.7  \\
\end{tabular}
\end{ruledtabular}
\end{table}
Owing to the complex structure of the spin lattice, the relation between the averaged couplings ($\bar J_1$, $\bar J_2$) and individual exchanges is more intricate than in $AA'$VO(PO$_4)_2$.\cite{tsirlin2009} Since there are two $J_i$ bonds yet single $J_i'$ and $J_i''$ bonds per lattice site, one writes $\bar J_i=(2J_i+J_i'+J_i'')/4$. Applying this expression to the DFT estimates in Table~\ref{tab:exchange}, we arrive at $\bar J_2=13.4$~K that is twice larger than the experimental $\bar J_2\simeq 6.6$~K. A similar two-fold overestimate is found in $AA'$VO(PO$_4)_2$ phosphates: compare $\bar J_2=18.4$~K from DFT to $\bar J_2=10.0$~K from the experiment in PbZnVO(PO$_4)_2$ (Ref.~\onlinecite{tsirlin2010}). This gives a strong support to the DFT results, and ascribes their apparent inaccuracy to systematic errors that do not alter qualitative trends.

The notable reduction in $\bar J_2$ (6.5~K in \navpof\ vs. $9-10$~K in PbZnVO(PO$_4)_2$ and similar phosphates) can be traced back to longer bonds comprising the superexchange pathways. The relevant parameters are \mbox{V--O} and O--O distances.\cite{tsirlin2010} The latter are edges of the PO$_4$ tetrahedra and stay nearly constant, whereas the former are flexible. In \navpof, the V--O distances exceed 2.0~\r A (Table~\ref{tab:distances}) and contrast with shorter ($1.95-2.0$~\r A) distances for leading AFM couplings in $AA'$VO(PO$_4)_2$ (Table~IV in Ref.~\onlinecite{tsirlin2009}). \navpof\ resembles Li$_2$VOSiO$_4$ in terms of the diagonal couplings ($\bar J_2=6.7$~K and 6.3~K, respectively), despite the fact that the PO$_4$ anion is smaller than SiO$_4$.

While the nature of the diagonal couplings in \navpof\ is readily comprehended, the evaluation of the NN couplings remains a challenge because of the computational inaccuracies of DFT. To make things even more complicated, the effects of the spatial anisotropy cannot be resolved by the available experimental data on \navpof. The orthorhombic distortion of the FSL\cite{[{For example: }][{}]starykh2004,*sindzingre2004,*bishop2008} favors a certain direction for the spin columns, stabilizes the columnar AFM state and, therefore, increases the saturation field $H_s$. When the experimental $H_s$ exceeds the prediction of the regular FSL model, their difference approximates the magnitude of the orthorhombic distortion for NN couplings.\cite{high-field} The spin lattice of \navpof, however, lacks any orthorhombic distortion and rather shows a complex tetragonal superstructure with three inequivalent couplings ($J_1,J_1'$, and $J_1''$). The couplings of each type run along both dimensions of the square lattice (Fig.~\ref{fig:gamma}) and do not stabilize the columnar AFM state, thereby no effect on $H_s$ should be expected. Presently, we are unable to give an experimental-based estimate of the spatial anisotropy in \navpof. As long as the DFT results are considered, appreciable anisotropy effects should be expected (Table~\ref{tab:exchange}). 

\begin{figure}
\includegraphics{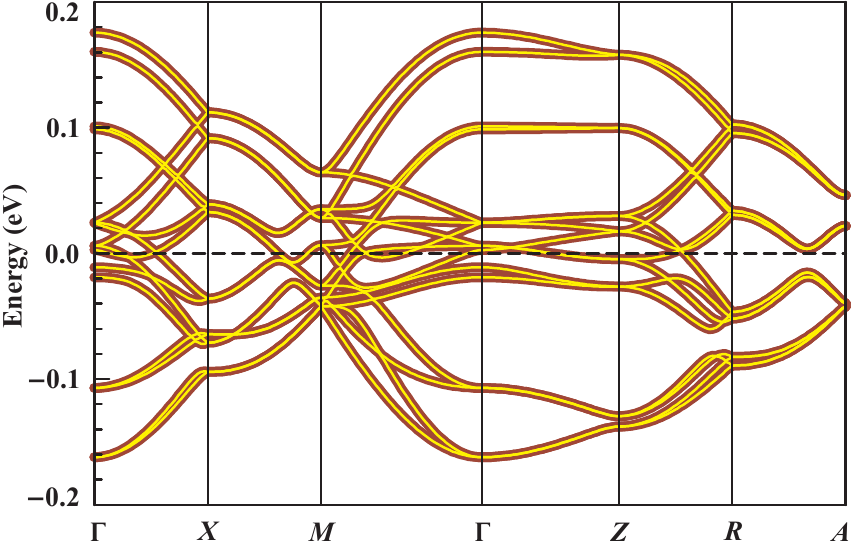}
\caption{\label{fig:band}
(Color online) Vanadium $d_{xy}$ bands at the Fermi level (zero energy): LDA band structure (thin light lines) and the fit of the tight-binding model (thick dark lines).
}
\end{figure}
The last important feature of \navpof\ is the sizable interlayer coupling $J_{\perp}$. It is related to dimers of octahedra sharing the fluorine atom (right panel of Fig.~\ref{fig:structure}), and couples each lattice site to one of the neighboring layers only. Since the magnetic saturation requires all spins to be parallel, the presence of the AFM interlayer coupling increases the saturation field. Eq.~\eqref{eq:hs} can be rewritten as
\begin{equation}
  \mu_0H_s=(2\bar J_1+4\bar J_2+J_{\perp})k_B/(g\mu_B)
\label{eq:hs-2}\end{equation}
to yield $J_{\perp}=1.1$~K, where we used $\bar J_1=-3.7$~K, $\bar J_2=6.6$~K, $\mu_0H_s=15.4$~T, and $g=1.95$. This \emph{experimental} estimate of $J_{\perp}$ compares well to the calculated value of 1.7~K (Table~\ref{tab:exchange}) and provides additional justification for the reliability of our DFT results.

The interlayer coupling in \navpof\ amounts to 20~\% of $\bar J_2$ and is larger than in any of the known FSL compounds (see Table~\ref{tab:comparison}). $J_{\perp}$ arises from the short V--V distance of 4.21~\r A, which is in fact the shortest V--V separation in the structure. There are comparable V--V separations of $4.45-4.55$~\r A in Li$_2$VOSiO$_4$ and Li$_2$VOGeO$_4$, but the lack of the bridging fluorine atom reduces the interlayer coupling to $J_{\perp}\simeq 0.2$~K and drives these compounds to a perfectly 2D regime.\cite{rosner2002,rosner2003} The $AA'$VO(PO$_4)_2$ phosphates show even weaker interlayer couplings $J_{\perp}\ll 0.1$~K\cite{tsirlin2010} owing to the V--V separations of $8.5-10.0$~\r A. The influence of the sizable interlayer coupling on the magnetism of \navpof\ is further discussed in Sec.~\ref{sec:disc-mag}.
\begin{figure*}
\includegraphics{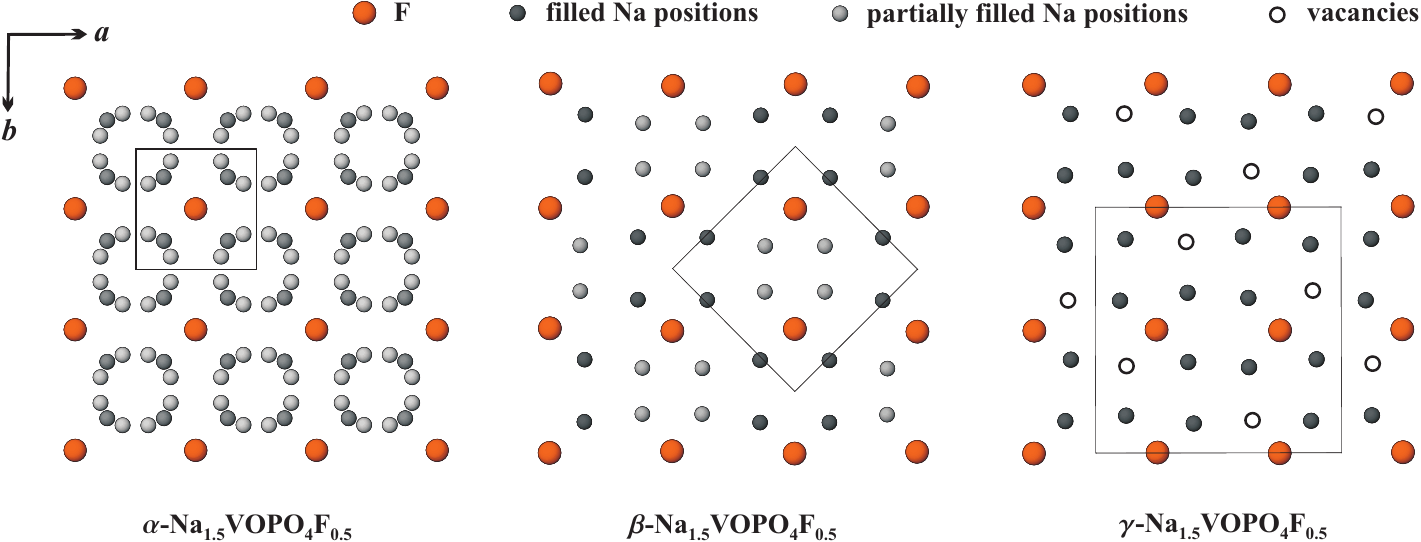}
\caption{\label{fig:Na}
(Color online) The arrangement of Na atoms in different polymorphs of \navpof. Dark spheres denote filled Na positions, light spheres are partially filled Na positions, and open circles represent vacancies.
}
\end{figure*}
\section{Discussion}
\label{sec:discussion}
\subsection{Crystal structure}
\label{sec:disc-cryst}
The crystal structure of \navpof, as reported in Refs.~\onlinecite{massa2002,sauvage2006}, is deceptive in its simplicity. It differs from our findings for \navpof\ prepared by high-temperature solid-state method. The $\alpha$-type structure, previously found at RT, is observed above 500~K only. The actual RT structure is marginally different, owing to the partial ordering of the Na atoms in the $\beta$-polymorph. Low-temperature studies reveal the formation of the complex superstructure in the $\gamma$-polymorph and the $\beta\leftrightarrow\gamma+\gamma'$ phase separation. The apparent differences in the RT structure are probably related to a variety of preparation procedures that could lead to slight deviations in the sample composition, especially the O/F ratio (see Sec.~\ref{sec:intro}). In our case, magnetization measurements combined with ESR and NMR local probes confirm the oxidation state of +4 for the vanadium atoms and ensure the ordered arrangement of O and F. Since our samples are synthesized from stoichiometric mixtures of the reactants, the \navpof\ composition is safely established.

The course of the structural transformations in \navpof\ is reminiscent of isotypic Na$_{1.5}M$PO$_4$F$_{1.5}$ fluorophosphates with $M^{+3}$ = Fe, Cr, V, Al, and Ga.\cite{structures1999} These compounds accommodate trivalent $M$ cations by substituting F for O1. This way, the $M$ cation attains a nearly regular octahedral coordination with fluorines in the axial positions. High-temperature structures are always of the $\alpha$-type, but low-temperature features are variable. When $M$ = Fe and V$^{+3}$, the RT structure is of the $\beta$-type, with $a=\sqrt2\,a_{\sub}$, $P4_2/mnm$ space group, and partially ordered Na atoms. The complete ordering of Na takes place in the orthorhombic $\gamma$-phase ($a\simeq b\simeq 2a_{\sub}$) below RT. By contrast, the fluorophosphates with $M$ = Cr, Al, and Ga drop directly to a $\gamma$-type RT polymorph having $a=2a_{\sub}$, $P4_2/mbc$ space group, and still disordered Na atoms. Below RT, these compounds presumably undergo another phase transition, but its nature has not been established. \navpof\ complements this welter of polymorphs and transitions by the ordered tetragonal $\gamma$-phase and the $\beta\leftrightarrow\gamma+\gamma'$ phase separation below RT.

Although the origin of the structural transformations in Na$_{1.5}M$PO$_4$F$_{1.5}$ is far from being clear, electronic effects should likely be ruled out. Indeed, magnetic (Cr) and non-magnetic (Al, Ga) trivalent cations give rise to the same sequence of the transitions.\cite{structures1999} In \navpof, the effect is even more tangible. Weak exchange couplings manifest themselves below $10-20$~K which is far below the first-order structural transition and phase separation around 250~K. In Fig.~\ref{fig:Na}, we plot the arrangement of Na atoms in different polymorphs of \navpof. The partial filling of the two multiple-site positions in the $\alpha$-polymorph is an indication of Na mobility. $\beta$-\navpof\ reveals clusters of ordered (Na1) and partially disordered (Na2) atoms. In $\gamma$-\navpof, we find a peculiar ordering pattern reminiscent of a $\frac14$-depleted square lattice. A similar pattern is observed in $\gamma$-Na$_{1.5}$FePO$_4$F$_{1.5}$, the only example of the complete cation ordering in this structure type.\cite{structures1999} The structural transformations/phase separation are possibly caused by electrostatic interactions between Na and F atoms together with an apparent difficulty of finding a symmetric configuration for the $\frac34$-filled ``square lattice'' of Na.

A notable feature of the phase separation in \navpof\ is the lack of any visible compositional changes. When electronic degrees of freedom are inactive, phase separation is typically driven by a size mismatch and results in different compositions of the two phases.\cite{[{For example: }][{}]NdLiTiO3-2007,*woodward2008} In \navpof, the fully ordered structure of the $\gamma$-phase implies the ideal \navpof\ composition and imposes the same composition for the $\alpha$-phase, as further confirmed by the unique oxidation state of +4 for vanadium. Neither ESR nor NMR reveal any differences between the two phases coexisting at low temperatures. Another surprising observation is the macroscopic scale of the separated phases: both $\gamma$- and $\gamma'$-polymorphs show narrow XRD reflections corresponding to well-defined crystallites rather than nanoscale domains. These unusual features suggest that a further study of the phase separation in \navpof\ could be insightful.
\subsection{Magnetic properties}
\label{sec:disc-mag}
\begin{table*}
\begin{minipage}{14cm}
\caption{\label{tab:comparison}
Comparison of FSL compounds with the columnar AFM ground state: exchange couplings $\bar J_1$ and $\bar J_2$ (in~K); thermodynamic energy scale $J_c=\sqrt{\bar J_1^2+\bar J_2^2}$ (in~K); the frustration ratio $\alpha=\bar J_2/\bar J_1$; the interlayer coupling $J_{\perp}$; and the magnetic ordering temperature $T_N$.
}
\begin{ruledtabular}
\begin{tabular}{crrrrrrc}\smallskip
  Compound & $\bar J_1$ & $\bar J_2$ & $\alpha$ & $J_c$ & $J_{\perp}/J_c$ & $T_N/J_c$ & Ref.     \\
  BaCdVO(PO$_4)_2$   & $-3.6$ & 3.2  & $-0.9$ & 4.8  & $<0.01$ & 0.21 & \onlinecite{nath2008}    \\\smallskip
  SrZnVO(PO$_4)_2$   & $-8.3$ & 8.9  & $-1.1$ & 12.2 & $<0.01$ & 0.22 & \onlinecite{kaul-thesis} \\
  BaZnVO(PO$_4)_2$   & $-5.0$ & 9.3  & $-1.9$ & 10.6 & $<0.01$ & 0.36 & \onlinecite{kaul-thesis} \\
  Pb$_2$VO(PO$_4)_2$ & $-5.2$ & 9.4  & $-1.8$ & 10.7 & $<0.01$ & 0.33 & \onlinecite{kaul2004}    \\
  PbZnVO(PO$_4)_2$   & $-5.2$ & 10.0 & $-1.9$ & 11.3 & $<0.01$ & 0.35 & \onlinecite{tsirlin2010} \\\smallskip
  \navpof            & $-3.7$ & 6.6  & $-1.8$ & 7.6  & 0.07    & 0.34 & This work                \\
  Li$_2$VOSiO$_4$    & 0.6    & 6.3  & +10.5  & 6.3  & 0.03    & 0.44 & \onlinecite{kaul-thesis} \\
  Li$_2$VOGeO$_4$    & 0.8    & 4.1  & +5.1   & 4.2  & 0.05    & 0.50 & \onlinecite{kaul-thesis} \\
\end{tabular}
\end{ruledtabular}
\end{minipage}
\end{table*}
Our detailed study of the magnetic properties and electronic structure confirms the assignment of \navpof\ to the quasi-2D FSL-type spin model. The ultimate goal of finding an ideal FSL with FM $J_1$ is, however, defeated by the complexity of structural transformations and low-temperature phase separation. The structural changes are not caused by the magnetic frustration (Sec.~\ref{sec:disc-cryst}), while the magnetism, in turn, is marginally influenced by the spin lattice distortion. Similar to $AA'$VO(PO$_4)_2$, thermodynamic properties conform to the regular FSL model. A closer and microscopic look, based on DFT, detects sizable deviations from the regular model, although experimental data give little information on such effects. In contrast to our earlier conjecture in Ref.~\onlinecite{high-field}, the underestimate of the saturation field should be ascribed to the interlayer coupling (Sec.~\ref{sec:band}). 

Despite seeming to be overcomplicated at first glance, the spin lattice comprising six inequivalent intralayer couplings (right panel of Fig.~\ref{fig:gamma}) expands the family of FSL-type models realized in inorganic compounds. The $AA'$VO(PO$_4)_2$ phosphates reveal a somewhat more simple, orthorhombic distortion of the FSL.\cite{tsirlin2009,[{For example: }][{}]starykh2004,*sindzingre2004,*bishop2008} By contrast, the tetragonal superstructure in $\gamma$-\navpof\ splits the square lattice into plaquettes (Fig.~\ref{fig:gamma}) and therefore relates to a different branch of theoretical results. The models of tetramerized square lattices, also known as square lattices with plaquette structure, exhibit a quantum phase transition between the long-range-ordered and gapped ground states but so far lack any prototype materials.\cite{[{For example: }][{}]singh1999,*kotov2001,*ueda2007,*albuquerque2008,*wenzel2008} A cation substitution/intercalation in \navpof\ could be a way to obtain further compounds of this type, although the possible problem of the phase separation should not be overlooked.

Apart from the spatial anisotropy, \navpof\ features an unusually strong interlayer coupling. In Table~\ref{tab:comparison}, we list some basic parameters of various FSL compounds with columnar AFM ground state: experimental values of $\bar J_1$ and $\bar J_2$, thermodynamic energy scale $J_c=\sqrt{\bar J_1^2+\bar J_2^2}$, DFT estimates of $J_{\perp}$, and experimental temperatures of the magnetic ordering ($T_N$). To account for the single interlayer coupling per lattice site in \navpof, we take one half of the experimental $J_{\perp}$. The comparison identifies the effects of the frustration and interlayer couplings on $T_N$. Based on $T_N/J_c$, the compounds can be divided into three groups: strongly frustrated ($\alpha\simeq -1$, $T_N/J_c\simeq 0.2$), moderately frustrated ($\alpha\simeq -2$, $T_N/J_c\simeq 0.35$), and weakly frustrated ($\alpha>5$, $T_N/J_c=0.45-0.50$) systems. An order of magnitude difference in $J_{\perp}/J_c$ between \navpof\ and PbZnVO(PO$_4)_2$ has no appreciable effect on $T_N$.

\begin{figure}
\includegraphics{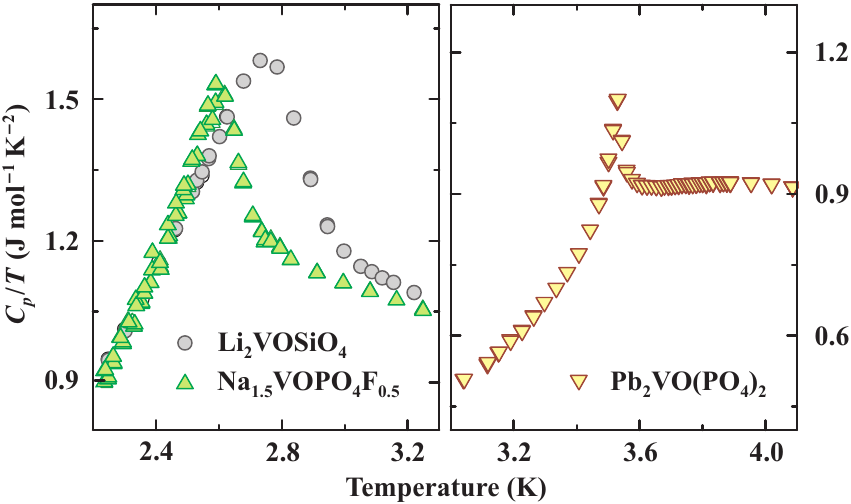}
\caption{\label{fig:anomalies}
(Color online) Zero-field specific heat at the magnetic ordering transition in \navpof\ (powder sample), Li$_2$VOSiO$_4$ (powder sample), and Pb$_2$VO(PO$_4)_2$ (single crystal).\cite{note8} The data for the last two compounds are from Ref.~\onlinecite{kaul-thesis}.
}
\end{figure}

Despite the weak influence on $T_N$, $J_{\perp}$ likely affects other features related to the magnetic ordering. In Fig.~\ref{fig:anomalies}, we compare specific-heat transition anomalies for Li$_2$VOSiO$_4$, \navpof, and Pb$_2$VO(PO$_4)_2$. The area under the $C_p/T$ curve is a measure of the transition entropy $S$.\cite{note4} Pb$_2$VO(PO$_4)_2$ shows the smallest $S$, below 0.01~J~mol$^{-1}$~K$^{-1}$. A similar frustration ratio of $\alpha\simeq -2$ leads to a much larger $S\simeq 0.065$~J~mol$^{-1}$~K$^{-1}$ in \navpof, while Li$_2$VOSiO$_4$ reveals an even broader anomaly with $S\simeq 0.11$~J~mol$^{-1}$~K$^{-1}$. Although numerical estimates of $S$ are tentative, the qualitative trend is robust and identifies the strong dependence of $S$ on $J_{\perp}$. The reduction in $J_{\perp}$ shifts the entropy to the broad maximum above $T_N$ and reduces the transition entropy, which determines the magnitude of the transition anomaly. At weak $J_{\perp}$, the anomaly ultimately transforms into a kink, especially for powder samples, see the data in Refs.~\onlinecite{kaul-thesis,nath2008,tsirlin2010}. A similar trend for the transition entropy depending on $J_{\perp}$ has been proposed in a theoretical study\cite{sengupta2003} of a simplified spin model with non-frustrated square lattices coupled by $J_{\perp}$. We mention that the 3D derivative of the FSL model has been addressed in recent theoretical studies focused on  ground-state properties.\cite{schmalfuss2006,*nunes2010,*majumdar2011,*oitmaa2011} Our results call for a further investigation of finite-temperature behavior: in particular, the evaluation of N\'eel temperature and temperature dependence of the specific heat.

Finally, \navpof\ can be a starting point to explore FSL-type systems with different values of the spin. Experimental work has been mostly restricted to the case of spin-$\frac12$,\cite{[{Note, however, recent work on insulating analogs of iron arsenides. For example: }][{}]fes,*cose} although theoretical results propose peculiar features of systems with larger spin.\cite{krueger2006,*moukouri2006,*bishop2008b,*jiang2009} The family of Na$_{1.5}M$PO$_4$F$_{1.5}$ fluorophosphates\cite{structures1999} should retain the basic FSL-type geometry with FM couplings between nearest neighbors and AFM couplings between next-nearest neighbors in the $ab$ plane. Trivalent $M$ cations represent spin-1 (V$^{+3}$), spin-$\frac32$ (Cr$^{+3}$), and even spin-$\frac52$ (Fe$^{+3}$) so that the effect of the spin value on the FSL model could be explored experimentally.

In summary, we have presented a detailed study of the crystal structure, magnetic behavior, and electronic structure of \navpof, a spin-$\frac12$ FSL-type compound. We observed essential structural changes, including the low-temperature phase separation, that prevent \navpof\ from being a simple model system. A complete experimental evaluation of the spin model for this compound is a formidable challenge. Despite this fact, \navpof\ is the only experimental example of the tetramerized (plaquette-structure) square lattice and bears a relation to a group of spin models which so far lacked any prototype materials. Our experimental data conform to the regular FSL model supplied with a sizable interlayer coupling $J_{\perp}$ of about 10~\% of the effective intralayer exchange. This interlayer coupling marginally affects the magnetic ordering temperature, yet the transition anomaly in the specific heat is notably increased. In general, \navpof\ and isostructural compounds establish several promising connections to FSL-type spin models. This should outweigh the apparent structural complexity of these materials and stimulate further experimental investigation.

\acknowledgments
We are grateful to ESRF for providing the beam time at ID31 and specifically acknowledge Andrew Fitch for his efficient support during the data collection. We would also like to thank Yurii Prots, Horst Borrmann, and Roman Shpanchenko for laboratory XRD measurements and Stefan Hoffmann for the DSC study. A.T. was supported by Alexander von Humboldt Foundation. This work was partially supported by the Deutsche Forschungsgemeinschaft (DFG) within the Transregional Collaborative
Research Center TRR 80 (Augsburg, Munich). Work at the Ames Laboratory was supported by the Department of Energy-Basic Energy Sciences under contract No. DE-AC02-07CH11358.


\begin{thebibliography}{80}
\bibitem [{not(){\natexlab{a}}}]{note2}%
  \BibitemOpen
  \bibinfo {note} {Throughout the manuscript, we list occupancies as fractions
  of the total occupancy for a given position.}\BibitemShut {Stop}%
\bibitem [{not(){\natexlab{b}}}]{note7}%
  \BibitemOpen
  \bibinfo {note} {The cell volume for $\beta$-polymorph slightly deviates from
  the linear behavior below 270~K. This might be a result of supercooling below
  the actual temperature of the $\beta\leftrightarrow\gamma+\gamma'$
  transition.}\BibitemShut {Stop}%
\bibitem [{not(){\natexlab{c}}}]{note3}%
  \BibitemOpen
  \bibinfo {note} {Owing to the presence of two phases with strongly
  overlapping reflections, we constrained ADPs for Na, P, and O,
  respectively.}\BibitemShut {Stop}%
\bibitem [{not(){\natexlab{d}}}]{note6}%
  \BibitemOpen
  \bibinfo {note} {Since $\chi=\chi_0+\chi_{\text{spin}}$, the difference
  between $\chi$ and $\chi_{\text{spin}}$ contributes to the chemical shift
  $K_0$ and does not alter the hyperfine coupling $A_{\text{hf}}$, see
  Eq.~\eqref{eq:shift}.}\BibitemShut {Stop}%
\bibitem [{not(){\natexlab{e}}}]{note1}%
  \BibitemOpen
  \bibinfo {note} {Similar to other V$^{+4}$ oxides, the FM contributions arise
  from the hoppings to $d_{yz}$ and $d_{xz}$ orbitals only. The hoppings to the
  $e_g$ states are negligible.}\BibitemShut {Stop}%
\bibitem [{not(){\natexlab{f}}}]{note4}%
  \BibitemOpen
  \bibinfo {note} {Experimental estimates of the transition entropy depend on
  the choice of the background, which is in general ambiguous. To get a rough
  estimate, we approximated the anomalies in Fig.~\ref{fig:anomalies} by
  triangles.}\BibitemShut {Stop}%
\end{thebibliography}
%
\newpage
\ 
\ 
\newpage
\renewcommand{\thefigure}{S\arabic{figure}}
\setcounter{figure}{0}
\begin{widetext}
\begin{center}
 \large
 \centerline{\textbf{Supplementary material}}
\end{center}
\bigskip
\begin{figure}[!h]
\includegraphics[width=11cm]{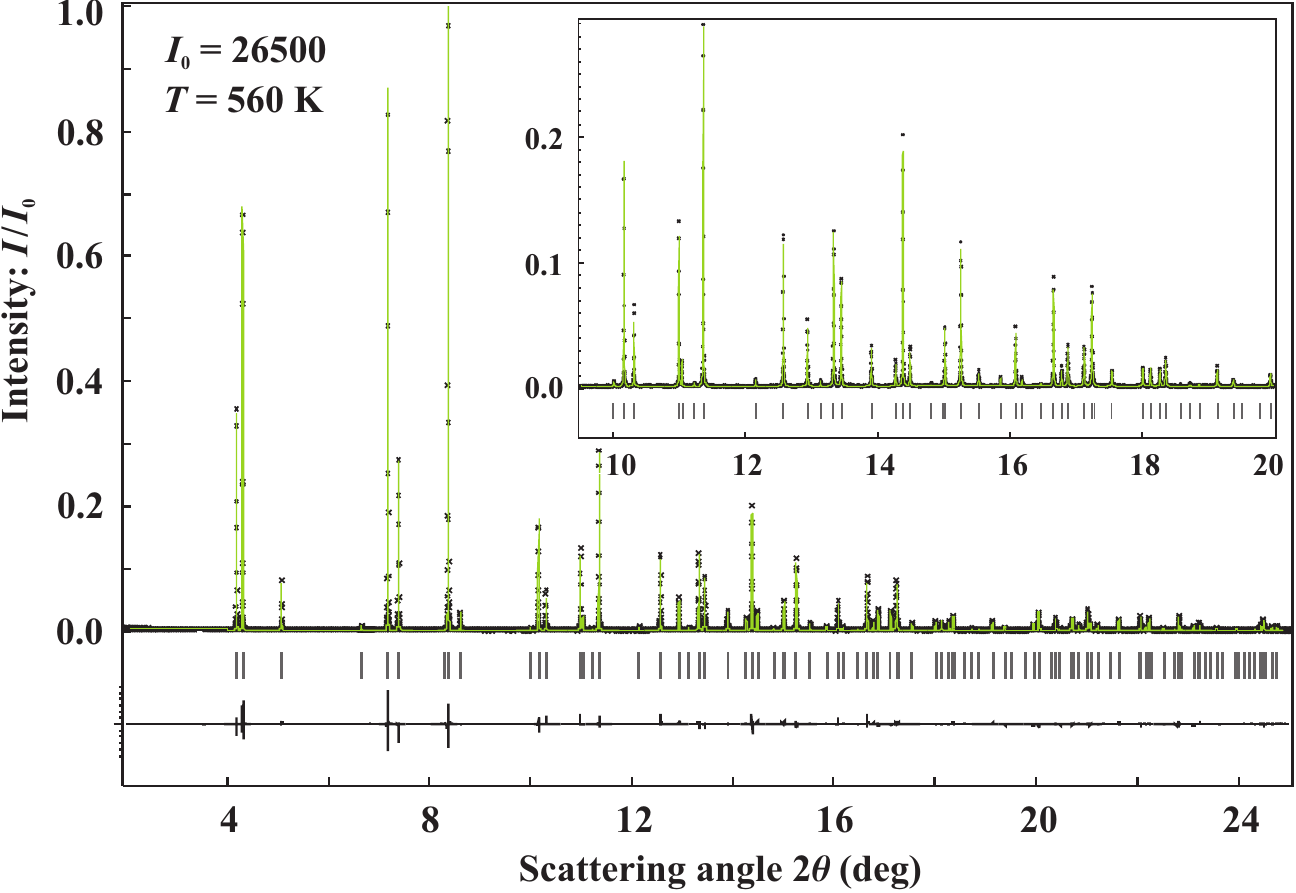}
\begin{minipage}{14cm}
\caption{\label{fig:s1}\normalsize
Structure refinement of $\alpha$-\navpof\ at 560~K: experimental (symbols), calculated (green line), and difference (black line) patterns. Ticks show reflection positions.
}
\end{minipage}
\end{figure}
\bigskip

\begin{figure}[!h]
\includegraphics[width=11cm]{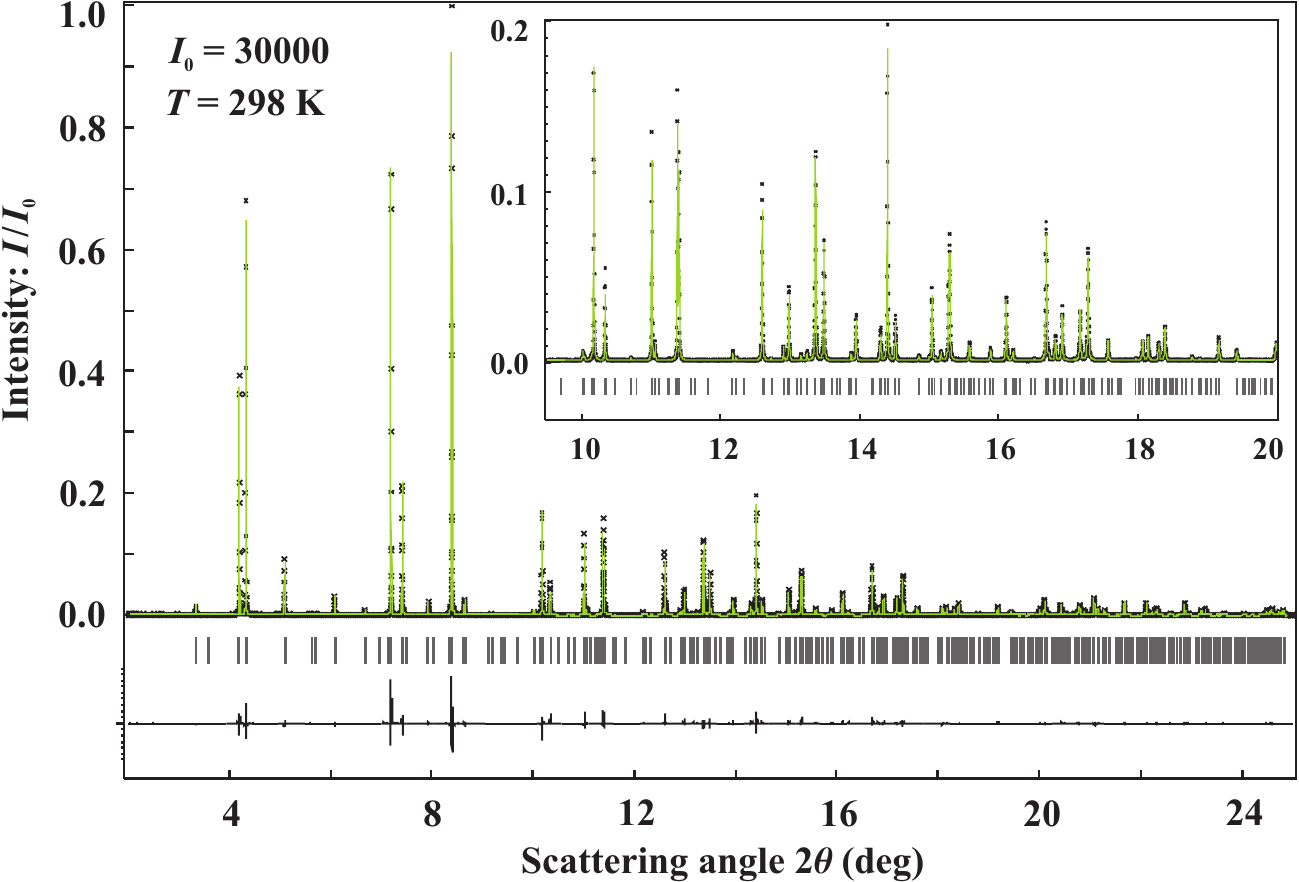}
\begin{minipage}{14cm}
\caption{\label{fig:s2}\normalsize
Structure refinement of $\beta$-\navpof\ at RT: experimental (symbols), calculated (green line), and difference (black line) patterns. Ticks show reflection positions.
}
\end{minipage}
\end{figure}
\bigskip

\begin{figure}[!h]
\includegraphics[width=11cm]{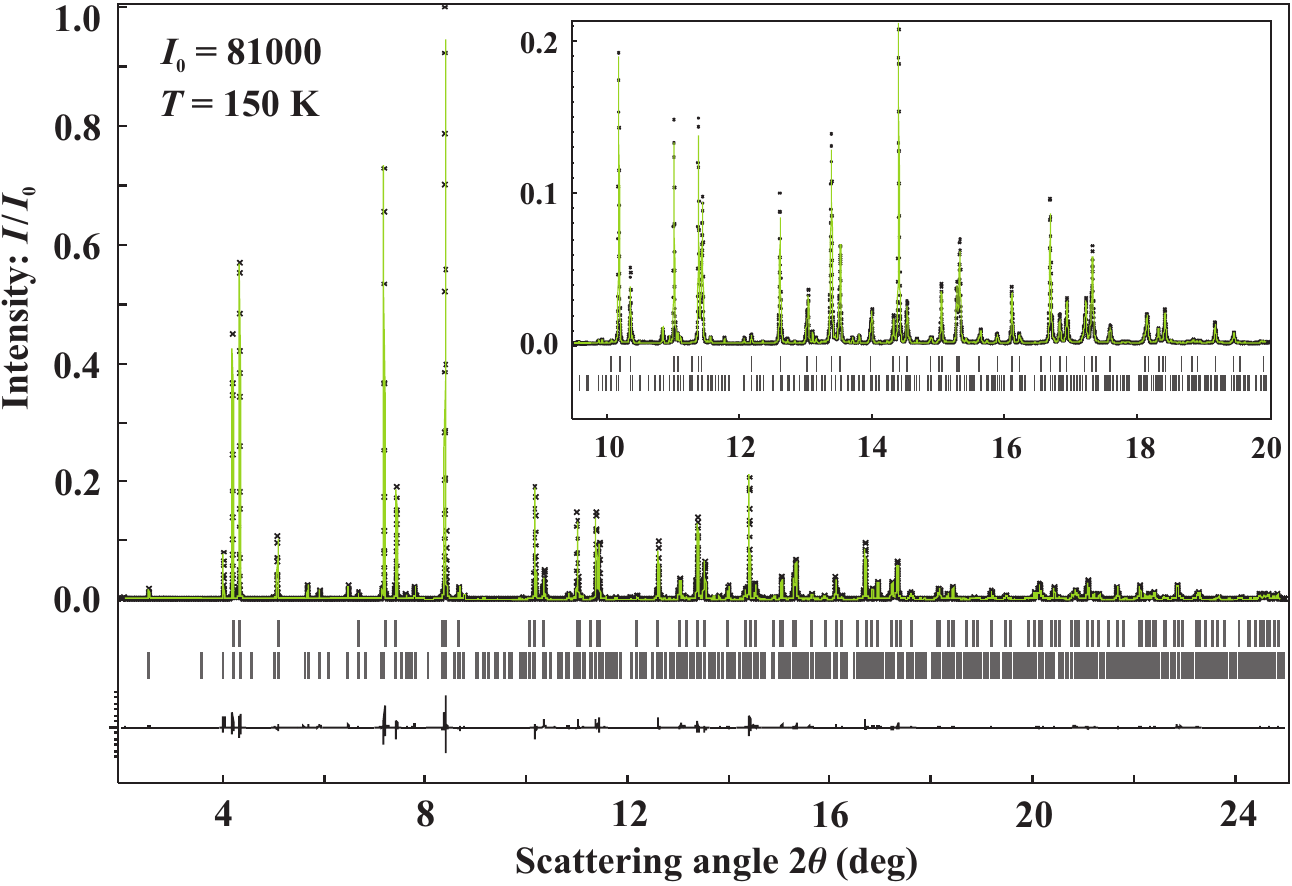}
\begin{minipage}{14cm}
\caption{\label{fig:s3}\normalsize
Structure refinement at 150~K: experimental (symbols), calculated (green line), and difference (black line) patterns. Upper and lower sets of ticks denote reflections of the $\gamma'$- and $\gamma$-polymorphs, respectively.
}
\end{minipage}
\end{figure}
\bigskip

\begin{figure}[!h]
\includegraphics[width=11cm]{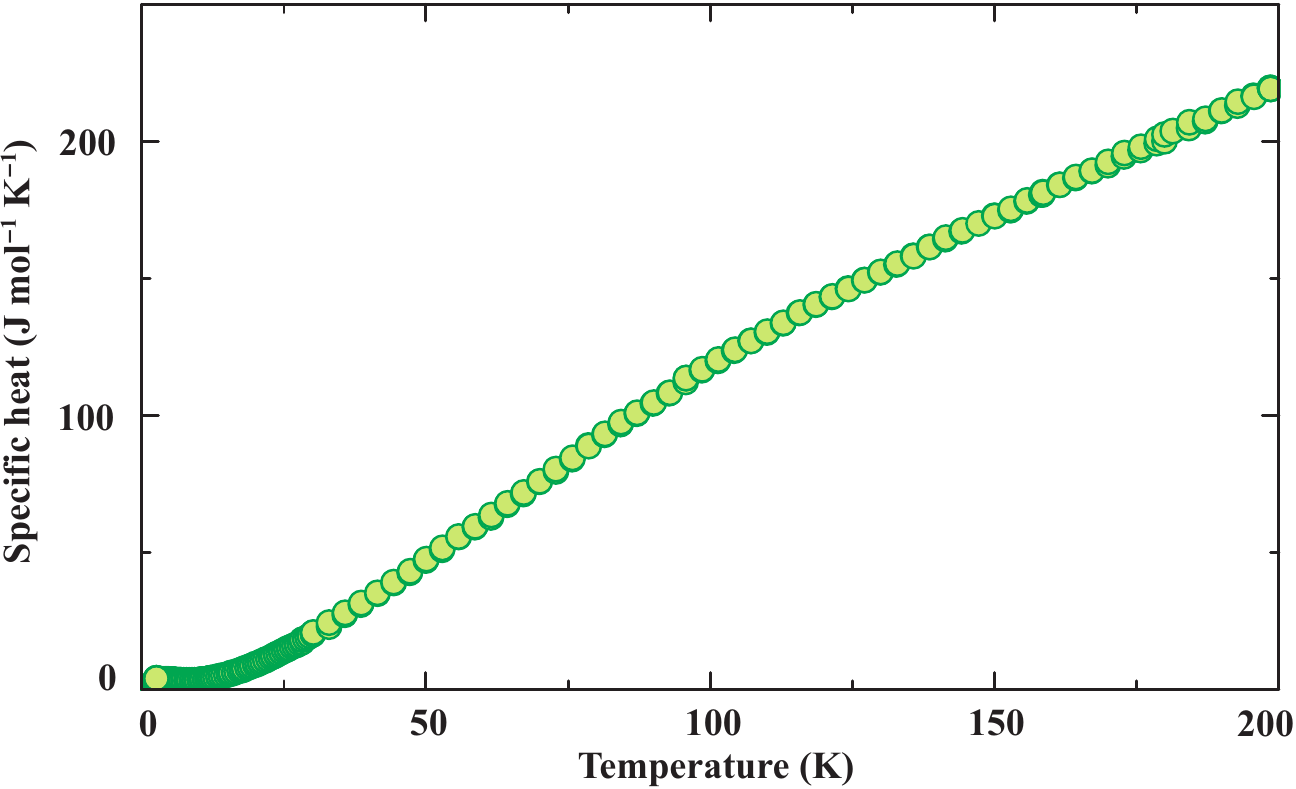}
\begin{minipage}{14cm}
\caption{\label{fig:s5}\normalsize
Specific heat of \navpof\ measured in zero magnetic field.
}
\end{minipage}
\end{figure}
\bigskip

\end{widetext}

\end{document}